\documentclass[aps,prl,longbibliography,twocolumn,superscriptaddress]{revtex4-2}

\usepackage{amsfonts}
\usepackage{amsmath}
\usepackage{graphicx}
\usepackage{dsfont}
\usepackage{diagbox}
\usepackage{array}
\usepackage{amssymb}
\usepackage{bbm}
\usepackage{float}
\usepackage{subfigure}
\usepackage{MnSymbol}
\usepackage{url}
\usepackage{times}
\usepackage{manfnt}
\usepackage{booktabs}
\usepackage{xcolor}
\definecolor{darkblue}{HTML}{004D6B}
\definecolor{darkred}{HTML}{8c1515}
\definecolor{darkgreen}{HTML}{006400}
\usepackage{hyperref}
\hypersetup{
    pdftitle={Robust teleportation of a surface code},
	colorlinks=true,
	urlcolor=darkred,
	citecolor=darkblue,
	linkcolor=darkred,
	breaklinks
}
\pagecolor{white}
\usepackage{ulem}
\usepackage{physics}
\usepackage{enumitem}
\usepackage{wasysym}

\begin{document}

\title{Robust teleportation of a surface code and cascade of topological quantum phase transitions}

\author{Finn Eckstein}
\affiliation{Institute for Theoretical Physics, University of Cologne, Z\"ulpicher Straße 77, 50937 Cologne, Germany}

\author{Bo Han}
\affiliation{Department of Condensed Matter Physics, Weizmann Institute of Science, Rehovot 7610001, Israel}

\author{Simon Trebst}
\affiliation{Institute for Theoretical Physics, University of Cologne, Z\"ulpicher Straße 77, 50937 Cologne, Germany}

\author{Guo-Yi Zhu}
\email{guoyizhu@hkust-gz.edu.cn}
\affiliation{Institute for Theoretical Physics, University of Cologne, Z\"ulpicher Straße 77, 50937 Cologne, Germany}
\affiliation{The Hong Kong University of Science and Technology (Guangzhou), Nansha, Guangzhou, 511400, Guangdong, China}

\date{\today}
\begin{abstract}
Teleportation is a facet where quantum measurements can act as a powerful resource in quantum physics, 
as local measurements allow to steer quantum information in a non-local way. While this has long been established for a single Bell pair, the teleportation of a many-qubit entangled state using non-maximally entangled resources presents a fundamentally different challenge. Here we investigate a tangible protocol for teleporting a long-range entangled surface code state using elementary Bell measurements and its stability in the presence of coherent errors that weaken the Bell entanglement. We relate the underlying threshold problem to the physics of anyon condensation under weak measurements and map it to a variant of the Ashkin-Teller model of statistical mechanics with Nishimori type disorder, which gives rise to a cascade of phase transitions. Tuning the angle of the local Bell measurements, we find a continuously varying threshold. Notably, the threshold moves to infinity for the $X+Z$ angle along the self-dual line -- indicating that  infinitesimally weak entanglement is sufficient in teleporting a self-dual topological surface code. Our teleportation protocol, which can be readily implemented in dynamically configurable Rydberg atom arrays, thereby gives guidance for a practical demonstration of the power of quantum measurements. 
\end{abstract}

\maketitle


The basis for {\sl fault-tolerant} quantum computation platforms are logical qubits that, built from many physical qubits, 
leverage long-range entanglement and topological protection to store quantum information \cite{Kitaev2003,Chen2010}.
One widely adopted blueprint for their implementation is the surface code \cite{Kitaev2003, Kitaev98surfacecode}, which like the 
toric code employs two commuting stabilizer measurements to induce a topological state of matter \footnote{Strictly speaking, when the code is defined on a planar (torus) geometry, it is called the surface (toric) code. Both surface and toric codes realize a $Z_2$ topologically ordered phase with the same anyon excitations and statistics.}.
Its fault-tolerance arises from the ability to perform quantum error correction based on the measurement outcomes of the
stabilizers (the so-called syndromes) and is embodied in a finite error threshold against incoherent noise 
such as non-deterministic Pauli errors~\cite{Preskill2002}. 
Going beyond a protected quantum memory, one of the most elementary building blocks for quantum information processing will be the {\sl teleportation} 
of a logical qubit, e.g.\ to spatially transfer quantum information non-locally such that it can be employed in a quantum circuit, 
akin to loading a classical bit into a processor register.
But while the teleportation of a single physical qubit is well studied both theoretically \cite{Bennett1993} and experimentally \cite{Bouwmeester1997,Hanson2014,Pan2017}, 
the teleportation of a logical state~\cite{Raussendorf2006, Raussendorf_2007, briegel2009measurement, Horsman2012surgery, Erhard2021latticesurgery, chou2018deterministic} supported through many qubits is a non-trivial challenge. 
This raises a fundamental question: how much entanglement is needed to teleport an {\it entire} topological surface code, including its long-range entanglement pattern that organizes the physical qubits~\cite{wen19choreographed}, and the logical qubits that they represent. 
If this task would be confined to the teleportation of a pristine many-qubit {\it wavefunction},
any source of decoherence would immediately make it unattainable. So the real question should be how one can preserve not the 
wavefunction but the associated  (topological) phase, 
such that the quantum information of a logical qubit  remains protected during teleportation even in the presence of decoherence.

In this manuscript, we address these questions by introducing a protocol for the teleportation of a many-qubit surface code state
and  demonstrate its ability to transfer a logical qubit. Introducing a source of coherent errors~\footnote{In our context, the coherent error is characterized by a tunable parameter that quantifies the imperfection of the unitary entangling gates in the quantum circuit away from the maximally entangled Clifford protocol.}
(by weakening the Bell measurements) we determine its robustness, threshold behavior, and optimal perfomance.
We recast these results in a many-body context by connecting the error threshold to an anyon condensation transition out of a 
topologically ordered quantum phase, which also gives an intuitive understanding to the remarkable robustness for certain Bell measurement angles. We provide additional analytical insights via a mapping of the problem to a classical Ashkin-Teller model~\cite{AshkinTeller43} with random (non-Hermitian) couplings whose statistical mechanics is reminiscent of Nishimori physics~\cite{Nishimori1981} in the random-bond Ising model (RBIM). On a conceptual level, our work goes beyond the widely studied phenomenology of Clifford decoherence and highlights the effect of {\sl non-Clifford} decoherence on the quantum many-body physics of a long-range entangled state, thereby shedding light on the stability of topological order in mixed states \cite{Fan23toriccode, Bao23errorfielddouble, Lee23criticalityunderdecoherence, Grover23decoherence, Mong24replicaTO}, making a connection to the physics of coherent errors \cite{bravyi2018correcting, Beri23toriccodecoherenterror, Beri24coherenterror, Iverson2020} and weak measurements \cite{NishimoriCat,chen2023realizing, Su23higherformsym, Garratt23measurement, Lee22, Alicea23measureIsing, zhu2023qubit, Schomerus2019, Ludwig2020, Altman2020weak, Schiro2021, Potter21review, Fisher2022reviewMIPT}, as well as generalized wave function deformations \cite{Fradkin04RK,Troyer10topocrit,Isakov2011,Zhu19}. With an eye towards experimental realization, we believe that configurable Rydberg atom arrays \cite{Browaeys2020,Lukin2021atom256,Lukin2023logical} can readily implement our protocol, controllably inject the coherent errors, and establish its fault-tolerance.


\begin{figure*}[t!] 
   \centering
   \includegraphics[width=\textwidth]{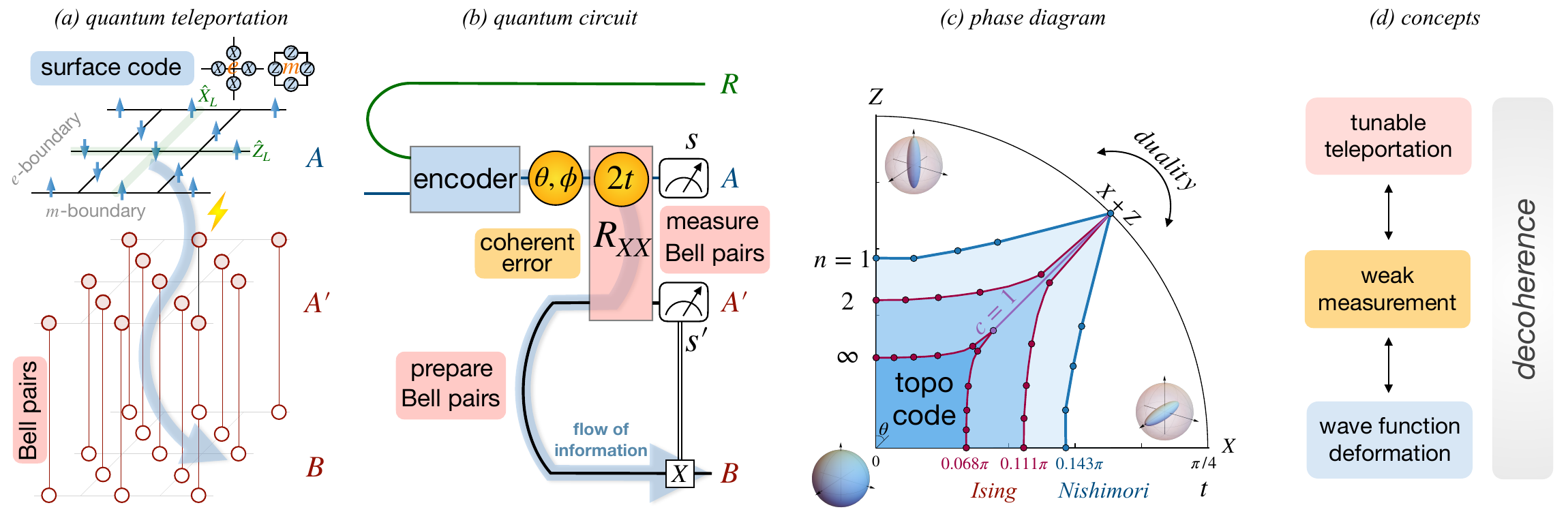} 
   \caption{
   {\bf Teleporting a logical qubit / surface code phase.}
   (a) Schematics of our teleporting protocol for $N$-qubit systems from $A$ to $B$. 
   Starting from a surface code state in $A$, it requires preparation of $N$ Bell pairs between auxiliary system $A'$ and $B$, 
   followed by subsequent Bell measurements for $A$ and $A'$. The shaded arrow indicates the flow of information. 
   (b) Quantum circuit. An encoder prepares a surface code state for $A$ qubits, whose logical qubit is maximally entangled with a reference qubit $R$ (via the green wire), while preserving the logical information. 
   $A'$ and $B$ are initialized in Bell pairs. The Bell measurement is performed by entangling $A$ and $A'$ via a unitary $R_{XX} $ gate, and subsequent measurements of $A$ and $A'$. A coherent error is introduced via the parameter $t$, which further depends on the Bell measurement angles $\theta,\phi$ induced by single qubit rotations on the Bloch sphere $U_{\theta,\phi} = \exp{-i Z\phi/2}\exp{-iY\theta/2}$ (orange circle). The shaded arrow corresponds to the information flow in (a) (if one remains below the threshold $t<t_c(\theta,\phi)$). 
    (c) Phase diagram in the $XZ$ plane with a cascade of transitions for the ensemble of post-measurement states. The duality is equivalent to a Hadamard transformation that swaps the $Z$ and $X$ axes, which yields a symmetric phase diagram.
   The origin is the fixed point surface code state for every possible measurement outcome.  The blue shaded regions stand for the topological surface code phases with protected code space maintaining 1-bit of $n$-th Rényi coherent information~\eqref{eq:purity}.
   The varying  shades of blue correspond to different fractions of the ensemble, separated by the red critical lines of varying replicas $n=2,\cdots,\infty$.
 The transition line (in blue) serves as the optimal phase boundary, beyond which the von Neumann coherent information ~\eqref{eq:coherentinfo} decays exponentially with the code distance.  
    The replica index serves as a ``lens" that can zoom into the more probable configurations of the measurement outcomes. Compared with the $n=1$ case that averages all configurations according to the Born's rule, the $n=2$ (outer red line) phase boundary falls into the Ising universality class except the 4-state Potts point at $t=\pi/4=\theta$. The $n=\infty$ (inner red line) phase boundary describes the post-selected $s=+1$ pure state among the ensemble, which is a cleanly deformed surface code state. At this phase boundary of $n=\infty$, the two Ising critical lines merge into a Kosterlitz-Thouless critical point at $t=\pi/8,\ \theta=\pi/4$, and opens up a gapless critical line for $\theta=\pi/4,\ t\geq \pi/8$. The dots are analytical or numerical data points. 
    (d) The phase diagram not only describes a teleportation protocol with coherent errors but also surface code under weak measurement, and wave function deformations resulting in topological transitions.   }
   \label{fig:protocol}
\end{figure*}


{\it Teleportation protocol}.--
The prototypical protocol for non-local quantum state teleportation \cite{Bennett1993} between two points in space 
requires three principal ingredients:
First, it needs Bell pairs to establish entanglement at arbitrary distance.
Second, quantum measurements enter with two consequences, steering the flow of quantum information from source to target and collapsing the initial wavefunction into a classical state. Finally, it requires a classical communication channel to transfer the measurement outcomes from source to target location, necessary to perform a round of corrections on the 
transferred quantum state (such as local qubit/basis rotations).

A schematic of our protocol for teleporting the quantum many-body state underlying a logical qubit is illustrated in Fig.~\ref{fig:protocol}(a), where we employ $N$ Bell pairs to steer the information between the spatially separated $N$-qubit systems $A$ (Alice) and $B$ (Bob). A more concise formulation is given in the quantum circuit of Fig.~\ref{fig:protocol}(b).
In the initialization stage, we prepare (i) a surface code state by an encoder for Alice's qubits $A$, and (ii) a bunch of $N$ Bell pairs, between the ancilla system $A'$ and Bob's qubits $B$. Concerning the degeneracy of the surface code states, we maximally entangle the logical state with a reference qubit $R$~\cite{Nielsen96coherentinfo}, such that when $R$ is traced out, the surface code becomes maximally mixed in the logical space. To perform the teleportation, we (spatially) align Alice's qubits $A$ with the ancilla qubits $A'$  one-to-one and perform a rotated Bell pair measurement for each pair, which is carried out by first applying a transversal unitary $R_{\rm XX} = e^{-i(\pi/4-t)X_1X_2}$ gates (Ising interaction evolution)~\cite{Lukin2023logical} and then projectively measuring both in the Pauli $Z$ basis. It is in this entangling step where we introduce a tuning knob (indicated by the orange circle), which allows to go to an imperfect entangling time (rotation) $0\leq t \leq \pi/4$, that weakens the entanglement between $A$ and $A'$. As a result, Alice effectively measures a {\it weakly entangled} pair: $(1-i\tan(\pi/4-t)X_1X_2)\ket{ss'}$, depending on the two measurement outcomes $s, s'=\pm1$. Here the weakly entangled Bell pair breaks the $SU(2)$ symmetry down to $U(1)$ with an axis dependence, where the axis can be rotated by a unitary $U_{\theta,\phi}$ from the Pauli $Z$ to an arbitrary directions on the Bloch sphere $\hat{\sigma}^{\theta,\phi} = \sin(\theta)\cos(\phi) X + \sin(\theta)\sin(\phi)Y + \cos(\theta) Z$, characterized by angles $\theta,\phi$. An example is $\ket{\nearrow\nearrow}-i\tan(\pi/4-t)\ket{\swarrow\swarrow} $ for $\theta=\pi/4, \phi=0$ rotating $Z$ to $Z+X$, with $\nearrow(\swarrow)$ denoting the positive (negative) eigenstate of $Z+X$. 
We find that the teleportation channel with such weakly entangled Bell pairs is given, up to a local unitary correction
~\cite{supplement}, by the following Kraus operator
\begin{equation}
M_s = \exp\left(\frac{\beta}{2}s\hat{\sigma}^{\theta,\phi}\right) / \sqrt{2\cosh(\beta)}  \ ,
\label{eq:measureop}
\end{equation}
where $s=\pm 1$ indicates Alice's measurement outcome and $\beta=\tanh^{-1}\sin(2t)$ characterizes the effective measurement strength. 
This channel can alternatively be interpreted as a weak measurement channel, due to the imperfect Bell measurement. 
On a conceptual level, these measurement gates implement a non-unitary, local {\sl wave function deformation} of the underlying topological state~\cite{Fradkin04RK, Troyer10topocrit, Isakov2011, Cirac2012, Zhu19, gmz20fibonacci, gmz20doublesemion, zhu22fracton}. 
Note also that $M_s \propto \left(1 + \tanh\frac{\beta}{2} s \hat{\sigma}^{\theta,\phi} \right)$ is a superposition of the identity and Pauli operators, which maps the density matrix onto $
M_s \rho M_s^\dag 
\propto \rho + \tanh^2\left(\frac{\beta}{2}\right) \hat{\sigma}^{\theta,\phi} \rho  \hat{\sigma}^{\theta,\phi}
+ s\tanh\left(\frac{\beta}{2}\right) \{\hat{\sigma}^{\theta,\phi} ,\rho\} \,,
$ where the off-diagonal terms render it distinct from the conventional Clifford Pauli errors.

Due to the quantum no-cloning theorem~\cite{Wootters1982nocloning}, 
the logical information either successfully flows to $B$ or leaks to the measurement outcomes of $A$. When $t=0$, a perfect Bell measurement is performed (which does not extract the logical information but rather propagates it across space) and the surface code state is successfully teleported to $B$, visualized as an information flow through the wire of the circuit from $A$ to $B$. 
When $t=\pi/4$, $A$ is decoupled from $B$, and the measurement collapses every qubit in the surface code, such that the information gets pumped out to the measurement outcomes of $A$ and cannot flow into $B$. 
When $0<t<\pi/4$, the variable strength can turn on and off the teleportation of the surface code, which will be shown to exhibit topological quantum phase transitions. However, the post-teleportation state depends on the measurement outcomes $\mathbf{s}$ (bitstring of $A$ qubits):
\begin{equation}
\ket{\Psi(\mathbf{s})} := \frac{ M_\mathbf{s} \ket{\Psi} }{\sqrt{P(\mathbf{s})}}\ ,
\label{eq:Psi}
\end{equation}
where $\ket{\Psi} :=\frac{1}{\sqrt{2}}( \ket{\psi_+}_B\ket{+}_R + \ket{\psi_-}_B\ket{-}_R)$ with $\psi_{+(-)}$ denoting the two degenerate surface code states, as eigenstates for the logical $\hat{X}_L$ operator. The normalization constant 
	$P(\mathbf{s}) = \bra{\Psi}M_\mathbf{s}^\dag M_\mathbf{s}^{\phantom \dag}\ket{\Psi} $ 
is the probability of measurement outcome according to Born's rule~\cite{NishimoriCat}. 
When all possible measurement outcomes are collected, with $A'$ traced out, the global state is a {\sl block-diagonal} mixed state
\begin{equation}
\rho_{RAB} = \sum_\mathbf{s} P(\mathbf{s})\ketbra{\Psi(\mathbf{s})} \otimes \ketbra{\mathbf{s}}_A \ .
\label{eq:rhoRAB}
\end{equation}
Under such effective decoherence induced by coherent error, the state remains topologically ordered iff it maintains a protected 2-dimensional code space, which means there exist 2 locally indistinguishable but global orthogonal states (akin to the degenerate ground states of a topological Hamiltonian~\cite{Kitaev2003}) in the thermodynamic limit of large code distances $d\to\infty$. The size of the protected code space can be detected by the coherent information~\cite{Nielsen96coherentinfo, Lloyd97coherentinfo, Gullans20scalabledecoder, Fan23toriccode, Mueller23coherentinfo}, which for $\rho_{RAB}$ is
\begin{equation}
I_c  
= S_{RA} - S_A
= S_{AB} - S_{RAB} 
= \sum_\mathbf{s} P(\mathbf{s}) S_B(\mathbf{s}) \ .
\label{eq:coherentinfo}
\end{equation}
One can put physical meaning to this formula in three different ways:
(i) $S_{RA} - S_A$ as a conditional entropy expresses the quantum information of $R$ being subtracted by the leakage into $A$, where $A$ plays a role analogous (but not identical) to the environment;
(ii) $S_{AB} - S_{RAB}$ expresses the quantum information that $B$ can decode with the assistance of classical information (measurement outcomes) from $A$;
(iii) $S_B(\mathbf{s})=S_R(\mathbf{s})$ is the von Neumann entropy of the {\it logical} qubit for each measurement outcome, which quantifies the size of uncorrupted quantum code space, whose average yields the coherent information. 
To calculate this quantity, note that the reduced density matrix of $R$ can be derived by projecting $M_{\bf s}^\dag M_{\bf s}^{\phantom \dag}$ onto the logical space of $B$~\cite{KnillLaflamme97}
\begin{equation}
\rho_R (\mathbf{s}) = \frac{1}{2P(\mathbf{s})} \left(
\begin{matrix}
P_{++}(\mathbf{s}) & P_{+-}(\mathbf{s}) \\
P_{+-}^*(\mathbf{s}) & P_{--}(\mathbf{s}) \\
\end{matrix}
\right)
\equiv \frac{1+\vec{\kappa}(\mathbf{s})\cdot\vec{\sigma}_R}{2} \ ,
\label{eq:rhoR}
\end{equation}
where $P_{\mu\nu} (\mathbf{s}):= \bra{\psi_\mu}M_\mathbf{s}^\dag M_\mathbf{s}^{\phantom\dag} \ket{\psi_\nu}$ is the overlap between two logical states being connected by the weak measurement operators, and $P(\mathbf{s}) = (P_{++}(\mathbf{s})  + P_{--}(\mathbf{s}) )/2$. 
Notably, Eq.~\eqref{eq:rhoR} can be interpreted as a qubit subject to a {\sl polarization field} vector $\vec{\kappa}$.
It expresses the precise logical error based on a fixed ``syndrome" $\mathbf{s}$
\begin{equation}
\mathcal{E}(\rho_L) = \sqrt{\rho_R(\mathbf{s})} \rho_L \sqrt{\rho_R(\mathbf{s})}\ ,
\label{eq:logicalerror}
\end{equation}
for any state $\rho_L$ in the logical space.
A finite $\kappa_{x(z)}$ compresses the logical Bloch sphere along the $X(Z)$ axis (Fig.~\ref{fig:protocol}), with density eigenvalues $(1\pm \kappa(\mathbf{s}))/2$.
For sufficiently large field strength, this shrinks the Bloch sphere to a {\sl classical} bit, which is read out by Alice -- indicating the teleportation phase transition
mapped out in Fig.~\ref{fig:protocol}(c).


{\it Topological degeneracy and anyon condensation}.--
To understand the general shape of the phase diagram in Fig.~\ref{fig:protocol}(c), it is helpful to relate the breakdown of teleportation
to the field-induced transition of the $Z_2$ gauge theory description underlying the surface code \cite{Wegner71duality,FradkinShenker79, Kogut79rmp,Trebst07toriccode,Tupitsyn2010, Dusuel2011, haegeman2015shadows, Zhu19, Nahum21selfdual, Verresen22higgs}. In this language, the surface code allows two types of elementary excitations: electric charge $e$ and magnetic flux $m$ particles, which due to their mutual semion statistics
are referred to as anyons. For the surface code open boundary condition (Fig.~\ref{fig:protocol}(a)), one can create two $e$ particles and separate them away from each other disappearing into the left and right $e$-boundaries~\cite{kitaev2012gappedboundary}, which transforms the surface code state $\psi_+$ into $\psi_-$, that are locally indistinguishable but globally orthogonal, yielding a 2-fold topological degeneracy -- this is the logical qubit space. Under small deformations~\eqref{eq:Psi}, the states $M_\mathbf{s} \ket{\psi_+}$ and $M_\mathbf{s} \ket{\psi_-}$ remain asymptotically orthogonal despite their anyon excitations starting to fluctuate. For large deformations, however, they become indistinguishable and the topological order breaks down. This phase transition is driven by the condensation of the anyons~\cite{Burnell18reviewanyoncondensation}. When an $e$-particle is condensed~\cite{Verresen22higgs}, $M_{\bf s}^\dag M_{\bf s}^{\phantom\dag}$ can map $\psi_+$ to $\psi_-$ leading to nonzero $\kappa_z$, which quantifies the $e$ condensation fraction. When an $m$-particle is condensed, the $e$-particles must be confined due to destructive interference with $m$. As a result, either $M_\mathbf{s} \ket{\psi_+}$ or $M_\mathbf{s} \ket{\psi_-}$ has exponentially decaying norm and is ill-defined, which is signalled by nonzero $\kappa_x$. In this aspect, the coherent information~\eqref{eq:coherentinfo} serves as a single order parameter that collects the deconfinement {\sl and} uncondensation contribution.


\begin{figure}[t] 
   \centering
   \includegraphics[width=\columnwidth]{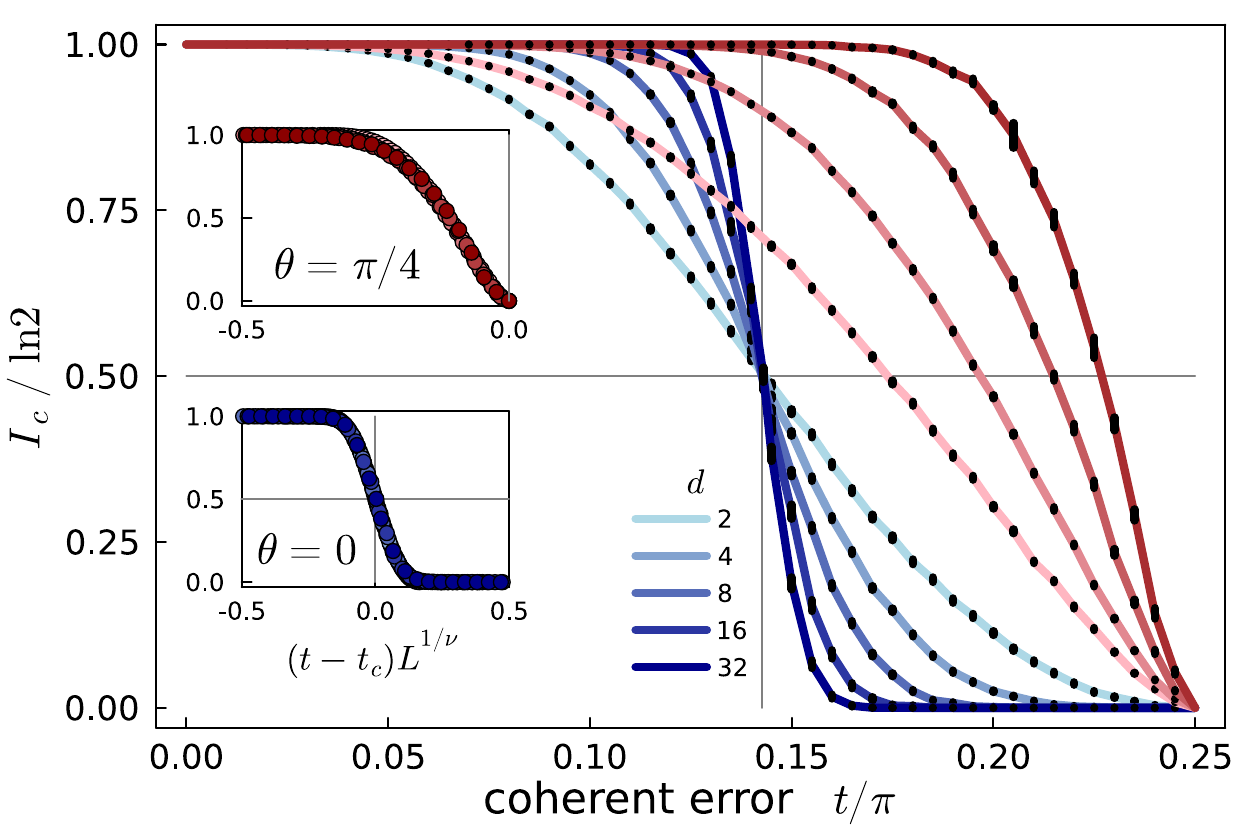}
   \caption{{\bf Coherent information and teleportation transition.}
   Shown are two sets of traces for $\theta=0$ ($Z$ direction) in blue, and $\theta=\pi/4$ ($X+Z$ self-dual direction) in red.
   Vertical gray lines indicate the thresholds, obtained from finite-size scaling analysis, with data collapses shown in the insets.
   (i) $\theta=0$: $t_c/\pi=0.143(1)$, $\nu=1.6(1)$, consistent with Nishimori criticality. 
   The gauge symmetry of the Nishimori line allows us to uncorrelate the disorder and perform random sampling, 
   where for each sample we perform tensor network contraction for the coherent information, 
   simulating code distances up to $d=32$ (1,985 qubits), averaged over 1000-10000 random samples. 
   (ii) $\theta=\pi/4$: $t_c/\pi=0.25(1)$, $\nu=1.8(1)$. 
   Without explicit gauge symmetry for the disorder ensemble, we Monte Carlo sample the disorder and subsequently 
   contract out the tensor network~\cite{NishimoriCat}. This hybrid approach allows us to simulate codes up to $d=16$ (481 qubits), averaged over 200-1000 random samples. 
   }
   \label{fig:numerics}
\end{figure}


{\it Mixed state phase diagram of Born average}.--
In Fig.~\ref{fig:protocol}(c), the thresholds / critical points are shown as the blue dots forming the blue line, inside which the entire inner blue shaded region stands for the topological  phase, where the average post-measurement state maintains the protected code space / topological degeneracy and constitutes a coherent superposition of loops~\cite{Kitaev2003}. 
In contrast, the states above the threshold decohere into a classical loop gas~\cite{Lee23criticalityunderdecoherence}, whose loops are indicated by the negative measurement outcomes $\{ s=-1 \}$. 
The two phases are separated by a phase transition whose precise location is obtained by a finite-size scaling collapse~\cite{malte_putz_2024_10712427} of the coherent information~\eqref{eq:coherentinfo}, which we computed using a hybrid Monte Carlo/tensor network technique for  shallow circuit sampling~\cite{NishimoriCat, itensor}, see Fig.~\ref{fig:numerics}.
Changing the Bloch angle $\theta$, the threshold $t_c$ is found to vary from a (Nishimori) transition  with threshold $t_c\approx 0.143\pi$~\cite{Nishimori1981, NishimoriCat} for $\theta=0$ off to $t_c = \pi/4$, ($\beta = \infty$, an ``$\infty$-threshold") for $\theta=\pi/4$, i.e.\ along the $X+Z$ Bloch projection.
To understand this, notice that for finite $\theta$ both $e$ and $m$ particles are fluctuating and compete with each other. As a result, it takes stronger deformation to achieve anyon condensation, which explains the enhancement of the threshold by deviating $\theta$ from $0$ or $\pi/2$. 
When $\theta=\pi/4$, a higher symmetry emerges as the state remains invariant under 
Hadamard transformations that swap $Z\leftrightarrow X$ for every physical qubit
-- this is the electric-magnetic self-duality~\cite{FradkinShenker79,Tupitsyn2010, Zhu19, Jian24dualitycode},
which along this line is respected not only by the pristine surface code ($t=0$) but also all deformations ($t>0$). 
As a consequence, the frustration from competing anyon condensation is strongest along this line and, as revealed in our calculations, 
pushes the threshold all the way to infinity. 
This implies a remarkable robustness of the teleportation protocol along this line.
In reverse, it means that teleportation of the topologically ordered many-qubit state between Alice and Bob is successful with only {\it infinitesimally weakly entangled pair resources} - the critically entangled pair we need to measure for a finite size system is $\ket{\nearrow\nearrow} + \mathcal{O}( d^{-1/\nu}) \ket{\swarrow\swarrow} $ (taking $s=s'=+1$ for example). 
For any experimental realization of surface code teleportation this is thus
the optimal angle. When recast in terms of weak measurement, this result tells us that a self-dual surface code is most robust 
against decoherence.



{\it Replicas and cascade of transitions}.--
To shed light on the ensemble of post-teleportation states, consider a Rényi variant of the coherent information
\begin{equation}
I_c^{(n)} =
\frac{1}{1-n}\ln \frac{\text{tr}( \rho_{RA}^n) }{ \text{tr}(\rho_A^n) }
=\frac{1}{1-n}\ln [ \text{tr}\rho_R(\mathbf{s})^n]_n
\ ,
\label{eq:purity}
\end{equation}
which on the r.h.s.\ is given as the logarithm of the average $n$-th order purity of the reference qubit. The $n$-replica average $[\cdots]_n := \sum_\mathbf{s}P(\mathbf{s})^n\langle \cdots\rangle/ \left(\sum_\mathbf{s}P(\mathbf{s})^n\right)$ can be viewed as linear average over $n$ replicas of the system carrying the {\sl same} disorder. Compared with the Born (1-replica) average discussed above, the $n$-replica average enhances the contribution of states with higher probability. 
The numerically computed 2-replica threshold is shown as the outer red line/dots in Fig.~\ref{fig:protocol}(c), which is generally smaller than the 1-replica threshold for varying angles -- with exception of the $\infty$-threshold at the self-dual angle $\theta=\pi/4$, which is preserved in the 2-replica system.
Unlike Eq.~\eqref{eq:coherentinfo}, the Rényi coherent information of the {\it mixed state} is distinct from the averaged {\it pure state} Rényi entropy~\footnote{The averaged pure state Rényi entropy is $S^{(n)} =\sum_\mathbf{s}P(\mathbf{s}) \left(\frac{1}{1-n} \ln \text{tr}\left(\rho^n\right)\right)$, in contrast to the Rényi coherent information $I_c^{(n)} = \frac{1}{1-n} \ln \frac{\sum_\mathbf{s}P(\mathbf{s})^n \text{tr}\left(\rho^n\right)}{\sum_\mathbf{s}P(\mathbf{s})^n}$ that averages the $n$-th order purity over the $n$-replicated disorder}. 
We conjecture that the 2-replica phase boundary will lower-bound the 1-replica and upper-bound the higher replica phase boundaries, for which a rigorous proof along the $Z$ and $X$ axes can be adapted from Ref.~\cite{Nishimori03replica}.


In the $\infty$-replica limit, $P(\mathbf{s})^n$ distills out only those configurations $\mathbf{s}$ that have the highest probability~\cite{Fan23toriccode, Nishimori03replica}, which usually post-selects the clean, frustration-free configuration $\mathbf{s}=+1$ so as to minimize the energy, in the language of statistical mechanics. This reduces Eq.~\eqref{eq:measureop} to a clean deformation operator, which can be treated analytically~\cite{Zhu19} and yields the inner phase boundaries marked by red/purple lines in Fig.~\ref{fig:protocol}(c). The $\infty$-replica generally exhibits the smallest threshold compared with $n=2$ and $n=1$, pointing to a {\sl cascade of phase transitions} where higher probability states generally have smaller thresholds. 
Note that along the self-dual line $\theta=\pi/4$, the $\infty$-replica stands out. It does not exhibit the same level of robustness
as found for $n=1,2$, but exhibits a finite threshold at $t_c=\pi/8$, beyond which the system exhibits a critical line 
(described by a $c=1$ conformal field theory with varying critical exponents) \cite{Zhu19}. 


\begin{figure}[t] 
   \centering
   \includegraphics[width=\columnwidth]{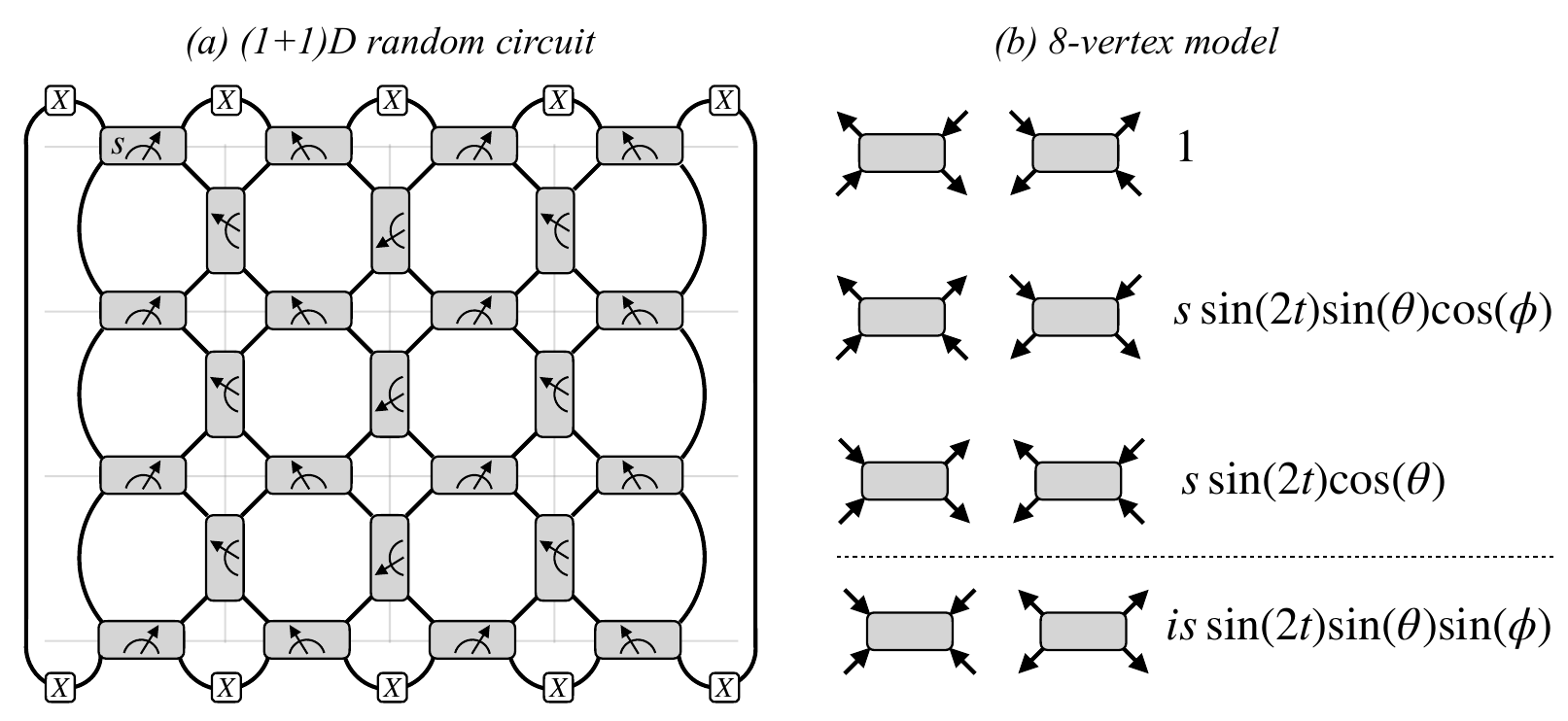} 
   \caption{{\bf Effective (1+1)D non-unitary circuit and 2D statistical model}. 
   (a) The surface code state under decoherence can be mapped to a non-unitary (1+1)D circuit, 
   by viewing one spatial dimension as fictitious ``time". 
   Each physical qubit in the surface code is mapped to a gate, 
   where the gate elements depend on the measurement outcome at the same location. 
   Their rectangle shape allows to distinguish the gates on the horizontal vs.\ verticals bonds. 
   When $\tan(\theta)\cos(\phi)=1$, the gate is self-dual: rotating the gate by 90 degree leaves it invariant
   and the network becomes invariant under vertex-plaquette duality, 
   consistent with the electric-magnetic self-duality of the surface code. 
   (b) The eight nonzero gate elements for the corresponding input and output spin configurations, which define a random 8-vertex model. The duality swaps the second and third row of vertices. 
   When $\phi=0$, the bottom two elements drop out, and the system reduces to a random 6-vertex model. 
   }
   \label{fig:2rep}
\end{figure}

{\it Statistical model}.--
To gain insight into the nature of the phase transition for generic $n$, we now proceed to map the quantum mechanical problem to a classical statistical model, akin to Ref.~\cite{NishimoriCat}'s approach (for Nishimori cat states). 
Following Born's rule, the probability function of the measurement outcome $P(\mathbf{s})$ is identical to the wave function amplitude, 
which here can be cast into a classical statistical model for two layers of spins (residing at the vertices) -- dual to the surface code wave function ket and bra, respectively.
As $P\equiv \sum_{\sigma,\tau}\exp(-\sum_{\langle ij\rangle}E_{ij})$, the pairwise spin interactions are 
\begin{equation}
\begin{split}
-E_{ij} &= 
J s_{ij} \frac{\sigma_i\sigma_j + \tau_i\tau_j}{2} 
+i\phi \frac{\sigma_i\sigma_j - \tau_i\tau_j}{2} \\
&+ (2K + i\pi \frac{1-s_{ij}}{2})\frac{\sigma_i\sigma_j\tau_i\tau_j-1}{2}
\,,
\end{split}
\label{eq:classicalmodel}
\end{equation}
with coupling strengths $\tanh (J)  = \sin(2t) \cos(\theta)$
and 
$e^{-2K}= \sinh(J) \tan(\theta) $ \cite{supplement}. 
This is an Ashkin-Teller model~\cite{AshkinTeller43} with generalized intra- and interlayer couplings.
First, there is a non-Hermitian term for finite Bloch angle $\phi$, 
i.e.\ when we consider a general deformation with Pauli $Y$ operators. We defer a discussion of this case to future work.
Second, the interlayer coupling exhibits random bond disorder introduced by the random measurement outcomes $s_{ij}$. 
For a general dictionary of the underlying quantum-classical correspondence, we refer to Tab.~\ref{tab:quantumclassicalcorrespondence}. The coherent information is then effectively determined by the boundary correlation~\cite{supplement}.

\begin{table}[t]
\centering
\caption{{\bf Quantum classical correspondence} between wavefunction and statistical model~\cite{Zhu19}. 
The ordering of classical spins $\sigma$ or their dual spins $\mu$ corresponds to the Higgs or confinement phase transition 
for the quantum wave function, respectively. 
}
\label{tab:quantumclassicalcorrespondence}
\begin{tabular}[t]{c   c  c}
\toprule
\,\,\,(2+0)D $\ket{\psi(\mathbf{s})}$\,\,\, & 2D Ashkin-Teller model & \,\,\,(1+1)D XXZ chain\,\,\, \\
\midrule
$e$ & $\sigma$ & \\
$m$ & $\mu$ & \\
$P_{--}/P_{++}$ & $\langle \sigma_0\tau_0\sigma_d\tau_d\rangle $ & $-\langle Z_0 Z_{2d+1}\rangle$ \\
$P_{+-}/P_{++}$ & $\langle \sigma_0\sigma_d\rangle $ & $\langle X_0X_{2d+1}\rangle$ \\
\bottomrule
\end{tabular}
\end{table}

In the classical model, the electric-magnetic duality of the quantum model turns into a Kramers-Wannier duality. This is most transparent in the 8-vertex representation of the Ashkin-Teller model, see Fig.~\ref{fig:2rep}(a) where it is equivalent to swapping the horizontal and vertical gates. For the Hermitian case ($\phi=0$), only 6-vertex configurations appear (Fig.~\ref{fig:2rep}b), and the transfer matrix of each slice describes a quantum XXZ chain of $2d+2$ spins with randomness. 
In such an XXZ representation, the logical operator becomes simply the correlation between the boundary spins (Tab.~\ref{tab:quantumclassicalcorrespondence}). 

These general quantum-classical mappings offer several merits. 
For one, the numerical exploration of the phase diagram is considerably more affordable in the Ashkin-Teller and particularly the XXZ 
representation. Second, by recasting the various thresholds/phase boundaries in terms of classical transitions we can infer their 
universality classes.
For the single-component $Z$ ($X$) transitions along the $\theta=0,\pi/2$ directions in our phase diagram, 
we can rigorously identify the 1-replica transition to be the Nishimori transition~\cite{Nishimori1981,NishimoriCat,Lee22,chen2023realizing}
of the 2D RBIM, 
while the 2-replica and $\infty$-replica transitions are non-random 2D Ising transitions \cite{Fan23toriccode,Zhu19}.
In the asymptotic limit $t\to\pi/4$ the $n$-replica model is, for all Bloch angles $\theta$,  an $S_{2n}$ permutation symmetric model (see SM~\cite{supplement}) with Kramers-Wannier duality, driven by the one-parameter coupling constant $J=\tanh^{-1}\cos(\theta)$. At $\theta\ll \pi/4$, every layer is ordered independently and $S_{2n}$ is spontaneously broken.
For the 2-replica case this leads us to conjecture that the Ising lines emanating from the $\theta=0$ transitions meet in a 
4-state Potts point (at $\theta=\pi/4,\ t_c=\pi/4$), akin to what happens in the $\infty$-replica case at finite threshold ($\theta=\pi/4,\ t_c=\pi/8$ where $J=2K$ gives rise to $S_4$ symmetric energetics $-K(\sigma_i\sigma_j+\tau_i\tau_j+\sigma_i\sigma_j\tau_i\tau_j)$.
This conjecture is corroborated by numerical simulations yielding a central charge estimate $c\approx 1$ from entanglement scaling \cite{supplement}.


{\it Discussion and outlook}.--
Zooming out, when the error preserves the self-duality of the surface code, the mutual frustration of anyon condensation points to a general guiding principle 
to dramatically enhance the code threshold. Beyond our work here, this is corroborated by the ``ultrahigh threshold" of the surface code under incoherent $Y$ noise~\cite{Flammia18Yerror, bonilla2021xzzx, claes2023tailored} and that under random projective Pauli measurements~\cite{Skinner2019, Barkeshli2021measuretoriccode, Vijay23Kitaev, Ippoliti23kitaev, Zhu23structuredVolumeLaw, Hsieh23toriccode, Yoshihito24measuretoriccode, Wei24measureToricCode} where self-duality is fulfilled on average~\cite{Mueller23surfacecodemeasure, Yoshida24monitoredcode}. The connection between percolation criticality for the latter case and the Nishimori transition reported here
is left for future study~\cite{Malte24}. 
In comparison with the self-dual {\it Hamiltonian phase diagram} of the 2D toric code or the $Z_2$ lattice gauge theory~\cite{FradkinShenker79, Trebst07toriccode, Tupitsyn2010, Dusuel2011, haegeman2015shadows, Nahum21selfdual, Verresen22higgs}, our surface code under imperfect teleportation or weak measurement exhibits a {\it wave function phase diagram} with distinct topology: while the ($e$ condensed) Higgs phase and the ($m$ condensed) confinement phase can be adiabatically connected beyond the finite extent of a first-order transition line in the Hamiltonian phase diagram, in our wave function phase diagram they are always separated by critical points (Fig.~\ref{fig:protocol}c). Besides, the Hamiltonian criticality is usually 3D conformal symmetric~\cite{Trebst07toriccode, Troyer10topocrit, Nahum21selfdual}, whereas the wave function criticality typically exhibits a dimensional reduction and belongs instead to 2D conformal criticality~\cite{ Fradkin04RK, Troyer10topocrit, Isakov2011, Zhu19}. 
This dimensional reduction is related to the fact that the quantum circuit of the whole protocol is {\it finite-depth} in its time dimension. 

To compare our  protocol to other teleportation schemes, it is useful to consider their required resources. Our protocol needs a set of $N$ independent Bell pairs, reflecting that it is a straightforward generalization of the standard teleportation protocol \cite{Bennett1993}  from few-body to many-body, where Bob's qubits are just a maximally mixed {\it product} state when others are traced out. This Bell-state approach can be compared with (i) the measurement-based quantum computation (MBQC) scheme~\cite{Raussendorf2001, Raussendorf2006, Raussendorf_2007, briegel2009measurement, briegel2009measurement} which prepares a joint {\it cluster state} for Alice and Bob or (ii) the transversal approach~\cite{Lukin2023logical} or the lattice surgery approach~\cite{Horsman2012surgery, Erhard2021latticesurgery} where Bob prepares a {\it long-range entangled} surface code beforehand. 
Because of its minimal request for Bob, our protocol is also highly suitable for a {\it highly non-local} transfer of a surface code, thus amenable for quantum communication or distributed topological quantum computing architectures.

\begin{figure}[b] 
   \centering
   \includegraphics[width=\columnwidth]{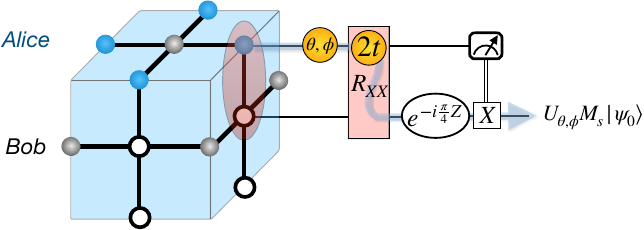} 
   \caption{{\bf Cluster-state based teleportation with coherent errors}. 
   Shown is a  3D Raussendorf lattice~\cite{Raussendorf2006} where the blue dots in the top layer are interpreted as Alice's qubits, 
   while the hollow circles in the second layer are Bob's qubits. 
   For convenience, we modify the conventional cluster state by a Hadamard transformation into $\prod_{\langle ij\rangle}e^{-i \frac{\pi}{4}X_iX_j}\ket{0}^{\otimes N}$. 
   The coherent errors $t$ weakens the otherwise maximally entangling gate $R_{XX}$ into $\exp{-i (\pi/4-t)XX}$. An additional unitary $U_{\theta,\phi}$ rotation is introduced before the entangling. The transferred wave function suffers from the same non-unitary error $M_s$ as in Eq.~\eqref{eq:measureop}. }
   \label{fig:fig5mbqc}
\end{figure}

Despite the microscopic difference of the various teleportation protocols, they share some commonalities on the level of their collective many-body physics. This includes, for instance, that -- independent of the type of  teleportation protocol -- the resource state prior to measurement exhibits symmetry-protected topological order \cite{Hong23teleport}.
%
Going further, one can even recast the existence of an error threshold and the nature of the ensuing phase transition between different protocols. Let us do this by restating our results for the Bell-state teleportation protocol in terms of the cluster-state based MBQC approach.
For our protocol we can show that if the resource Bell pairs are imperfectly prepared, e.g.\ by weakly entangling gates,  this will not immediately impede teleportation~\cite{supplement}, as Bell measurement and Bell preparation play dual roles~\footnote{
For instance, along the self-dual direction, the critical entanglement entropy between $A'$ and $B$ scales as $\mathcal{O}(d^{2-1/\nu})$, contributed by $\mathcal{O}(d^2)$ number of weakly entangled pairs, each of which has power-law scaling entanglement entropy $\mathcal{O}(d^{-1/\nu})$. 
}
in fostering a quantum channel across space~\cite{Bennett1993}. This is what ultimately gives rise to the phase diagram of Fig.~\ref{fig:protocol}(c). 
Notably, one can cast this phase diagram for the imperfectly prepared Bell-state teleportation protoocal also as a statement about the fault-tolerance of the imperfectly prepared cluster-state based MBQC approach~\cite{Raussendorf2006, Raussendorf_2007, briegel2009measurement} in the presence of coherent errors. To see this, consider two layers of the 3D Raussendorf cluster state~\cite{Raussendorf_2007}, as illustrated in Fig.~\ref{fig:fig5mbqc}, where a measurement by Alice (on the top layer) allows the surface code to be transferred to Bob (on the second layer). When one now considers a scenario where the cluster state is prepared in an imperfect manner by non-maximally entangling gates, this not only reduces the entanglement of the resource state (akin to what we have considered for the Bell state approach), but one finds that the effective error  takes precisely the form of the Kraus operator in Eq.~\eqref{eq:measureop} (see the Appendix for a detailed derivation). Consequently, the ensuing thresholds for successful transfer of the logical state in the cluster-state based approach is described in one-to-one correspondence by our phase diagram in Fig.~\ref{fig:protocol}(c).


\begin{figure}[t] 
   \centering
   \includegraphics[width=\columnwidth]{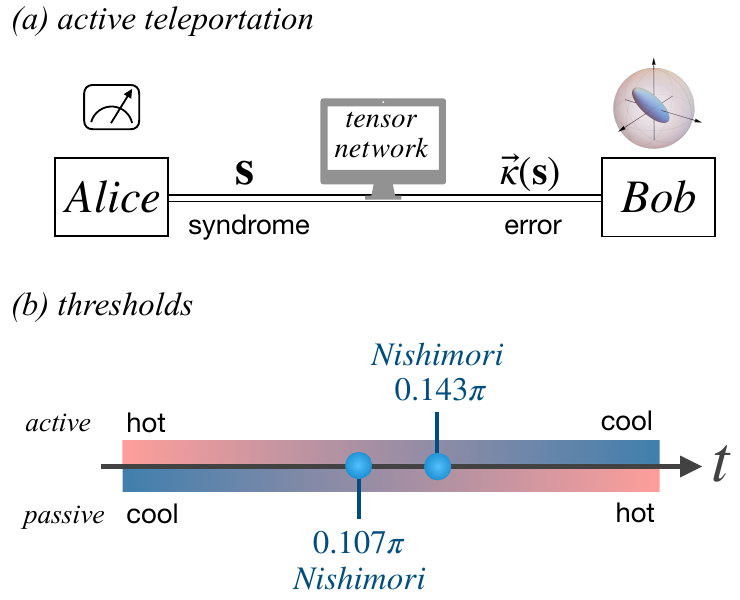} 
   \caption{{\bf Active vs.\ passive teleportation.}
   (a) Active teleportation: Alice sends her measurement outcome $\mathbf{s}$ to a classical computer that runs the tensor network calculation to deduce the error on the logical qubit. The result is fed to Bob such that Bob can correct the code accordingly. 
   (b) Threshold comparison between active and passive teleportation protocols for $\theta=0,\ \phi=0$.
   }
   \label{fig:decoder}
\end{figure}
Finally, let us mention that an experimental realization of {\sl many-qubit} teleportation~\cite{Alicea24teleportation} can benefit from implementing an error correction scheme. For this one can follow two different paths, which we dub ``active" versus ``passive" teleportation:
To decode Bob's deformed surface code one can use Alice's information $\mathbf{s}$  as syndrome -- this is what we call ``active teleportation". Here one can then implement a  tensor-network decoder~\cite{Bravyi14tensornetworkdecoder}, i.e.\ one can run our tensor network calculation for the random circuit in Fig.~\ref{fig:2rep} to compute the precise logical error in Eq.~\eqref{eq:logicalerror}, which can then be inverted by feedback operation or post-processing (Fig.~\ref{fig:decoder}a). The teleportation transition can be alternatively diagnosed by investigating whether Alice learns half of the logical information from $\mathbf{s}$ via a decoder~\cite{supplement}, which causes the logical qubit to collapse to a logical classical bit, similar to the scalable decoder for measurement-induced entanglement phase transitions~\cite{Gullans20scalabledecoder} or learnability transition~\cite{Vasseur22learnability}. 

Alternatively, if we do {\it not} use Alice's information $\mathbf{s}$ as syndrome -- a scenario which we dub ``passive teleportation", different pure states $\Psi(\mathbf{s})$ mix together, equivalent to an incoherent Pauli noise channel (where the off-diagonal terms are erased akin to the Pauli twirling for coherent errors~\cite{bravyi2018correcting, Beri23toriccodecoherenterror}):
$
\mathcal{N}(\rho) = (1-p) \rho + p \hat{\sigma}^{\theta,\phi} \rho \hat{\sigma}^{\theta,\phi}$ with $p=\sin^2(t)$. 
Such ``ignorance" of the measurement outcomes is generally expected to lead to smaller thresholds. 
Take $\theta=0 ,\ \phi=0$ for instance (Fig.~\ref{fig:decoder}b): {\sl with} Alice's knowledge, the RBIM with disorder probability $p = \sin^2(\pi/4-t)$ gives a threshold at $t_c\approx0.143\pi$; in contrast, {\sl without} Alice's knowledge the mixture is subjected to dephasing noise~\cite{Preskill2002} described by a RBIM with disorder probability $\sin^2(t)$, which yields a threshold of $t_c\approx 0.107\pi$ instead. 
The cascade of transitions for such passive teleportation is thus an example of the recently explored notion of {\sl mixed-state topological order} transitions~\cite{Bao23errorfielddouble, Fan23toriccode, Grover23decoherence, Mong24replicaTO} driven by anyon condensation in the double-Hilbert space -- as opposed
to anyon condensation in the single-Hilbert space for active teleportation. 
By tuning the ``degree of ignorance", one can join the active  and the passive teleportation transitions unifying the two anyon condensation mechanisms, which we leave to future investigations.


\begin{acknowledgments}
{\it Note added.--}
Shortly after our work was posted on arXiv, a preprint~\cite{Grover24selfdual} reports the toric code under self-dual Pauli noise, which is equivalent to our passive teleportation protocol, where they show a qualitatively similar phase diagram with ours for the 2-replica model; a preprint~\cite{Hart2024} has reported the experimental realization of a deformed surface code used in a parity game to demonstrate quantum advantage, whose phase transition is identical to the $Z$ axis in our phase diagram; a preprint~\cite{Ryan24logicalteleport} reports logical teleportation using trapped ions, where Alice and Bob both have prepared long-range entangled states before the teleportation protocol. 

\textit{Acknowledgments.--} 
GYZ would like to especially thank Jason Alicea for inspiring discussions on entanglement swap and teleportation during the {\it Majorana quo vadis} workshop in Cologne and a subsequent visit to Caltech.
We thank Markus Müller, Sara Murciano, Ze-min Huang, Luis Colmenarez, Sebastian Diehl, Samuel Garratt, Michael Buchhold, Ruben Verresen, Nathanan Tantivasadakarn, Michael Gullans, Dominik Hangleiter, Malte Pütz, Andreas Ludwig, and Hidetoshi Nishimori for helpful discussions. 
The Cologne group acknowledges partial funding from the Deutsche Forschungsgemeinschaft (DFG, German Research Foundation)
under Germany's Excellence Strategy -- Cluster of Excellence Matter and Light for Quantum Computing (ML4Q) EXC 2004/1 -- 390534769 
and within the CRC network TR 183 (Project Grant No.~277101999) as part of subproject B01. 
BH acknowledges support from CRC 183, a Koshland Postdoc Fellowship and the hospitality of the Institute for Theoretical Physics during an extended stay in Cologne.
The numerical simulations were performed on the JUWELS cluster at the Forschungszentrum Juelich. 
GYZ acknowledges the support by the start-up fund from HKUST(GZ). \\
\end{acknowledgments}


{\it Data availability}.-- 
The numerical data shown in the figures and the data for sweeping the phase diagram is available on Zenodo~\cite{zenodo_teleportation}.


\bibliography{measurements}


\clearpage
\appendix

\setcounter{secnumdepth}{2}

\clearpage
\section{Supplementary data for self-dual line $\mathbf{X+Z}$}

\subsection{2-replica model}
\begin{figure}[htbp] 
   \centering
   \includegraphics[width=\columnwidth]{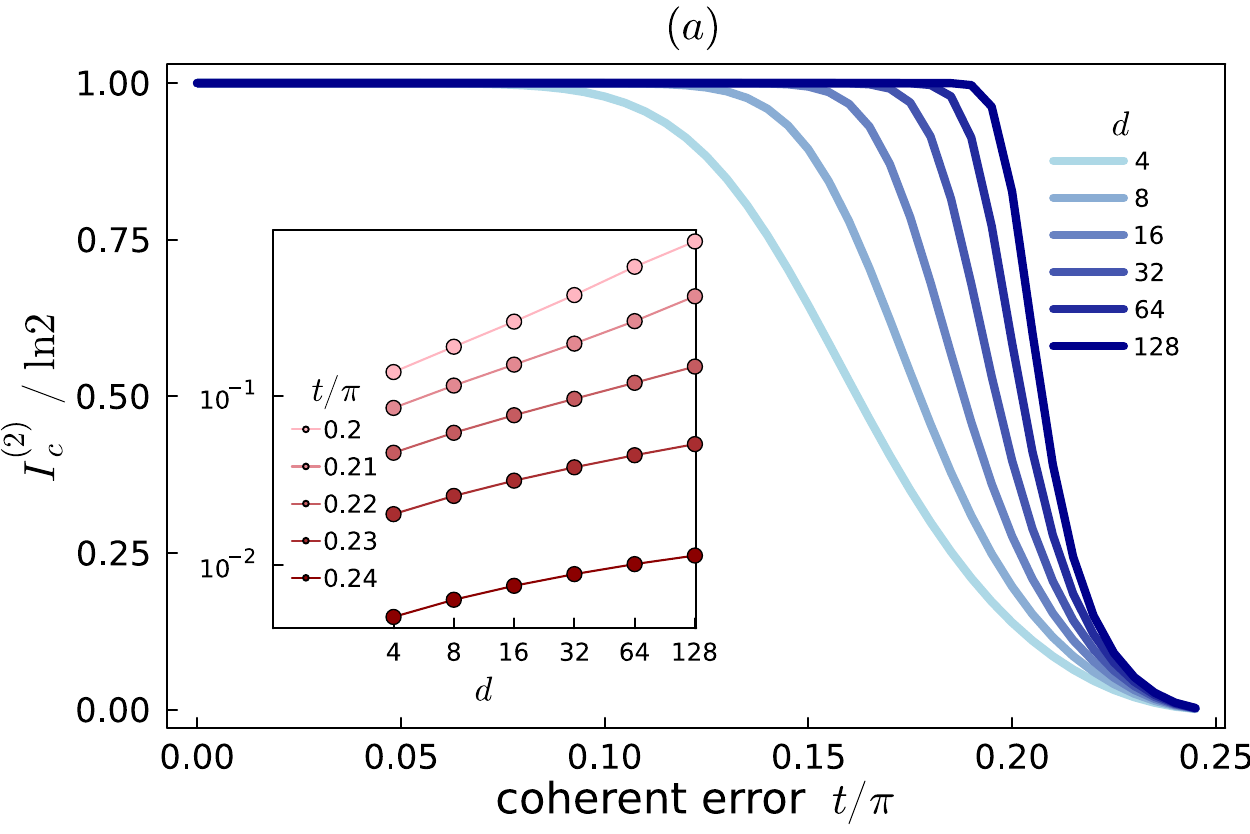} 
   \includegraphics[width=\columnwidth]{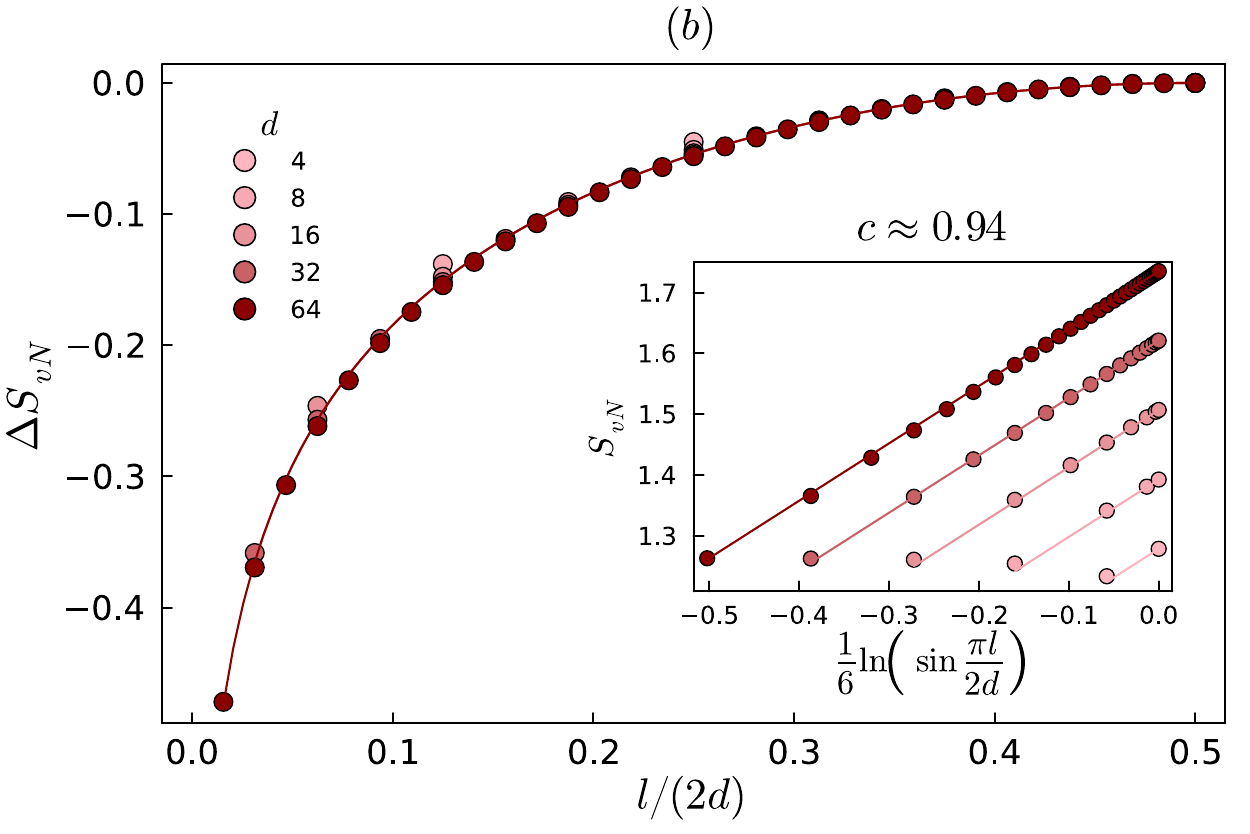} 
   \caption{{\bf 2-replica model along self-dual line  $\mathbf{X+Z}$}.
   (a) {Second Rényi coherent information sweeps of the phase diagram}. 
   	MPS virtual bonds are truncated by dropping density eigenvalues $\leq 10^{-10}$, 
	and bounded by maximal bond dimension $\chi=256$.
   (b) {Criticality at $t\to\pi/4$ by von Neumann entanglement entropy} of the (1+1)D boundary MPS ($\chi=1024$).
	Data points are shows as dots, lines indicate a fit that reveals a central charge $c\approx 0.94$.    }
   \label{fig:2repselfdual}
\end{figure}

Here we investigate the 2-replica XXZ quantum spin chains. Along the self-dual $X+Z$ line, $\sinh(J)=e^{-2K}$, which simplifies each local transfer matrix into
\begin{equation}
\begin{split}
T^{(2)}  &\propto \left(\frac{1-Z_{j}Z_{j+1}}{2}\right)^{\otimes 2}+ \frac{\sin^2(2t)}{2} \left(\frac{1+\vec{\sigma}_{j}\cdot \vec{\sigma}_{j+1}}{2}\right)^{\otimes 2} \ ,
\end{split}
\end{equation}
for both horizontal and vertical gates (Fig.~3 in main text). Note that $(1+\vec{\sigma_i}\cdot\vec{\sigma_j})/2$ is a swap operator, whose eigenstates are the spin singlet and triplets with eigenvalues $\mp 1$, respectively. As shown in Fig.~\ref{fig:2repselfdual}, the 2nd Rényi coherent information is computed by the 2-replica average of the purity $I_c^{(2)} = -\ln [\text{tr}\rho_R^2]_2= \ln 2-\ln (1+[\kappa^2]_2)$, as a function of the two point correlation of the 2-replicated XXZ chain
\begin{equation}
\begin{split}
[\kappa^2]_2 
=&
\frac{
\sum_\mathbf{s} (P_{++} - P_{--})^2 + 4|P_{+-}|^2}
{
\sum_\mathbf{s} (P_{++} + P_{--})^2
}\\
=&
\frac{
\langle (1-\sigma_0\tau_0\sigma_d\tau_d)^{\otimes 2}\rangle
+
4\langle (\sigma_0\sigma_d)^{\otimes 2}\rangle
}
{
\langle (1+\sigma_0\tau_0\sigma_d\tau_d)^{\otimes 2}\rangle
}\\
=&
\frac{
\langle (1+Z_0 Z_{2d+1})^{\otimes 2}\rangle
+
4\langle (X_0 X_{2d+1})^{\otimes 2}\rangle
}
{
\langle (1-Z_0 Z_{2d+1})^{\otimes 2}\rangle
}  \ ,
\end{split}
\end{equation}
where from the second line to the third line we use the operator map~\eqref{eq:mapAT2XXZ} and conserved quantity $\prod X = +1,\ \prod Z = (-1)^{d+1}$ of the spin chain for both copies. 
At asymptotic limit $t\to\pi/4$, the entanglement entropy of the boundary state conforms to the Calabrese-Cardy formula $S_{vN} = \frac{c}{6} \ln\sin(\pi l/(2d))+\cdots$, consistent with the $c=1$ CFT, which combined with the $S_4$ permutation symmetry points to the 4-state Potts CFT. 
Numerical details for Fig.~\ref{fig:2repselfdual}b: we elongate the depth of the circuit from $d$ rows to $4d$ rows to ensure steady boundary state, and drop the leftmost and rightmost dangling qubit leaving a chain of $2d$ qubits under the conventional open boundary condition. The entanglement cut is performed on the even cuts through the vertical gates. 

\subsection{$\infty$-replica model (post-selection)}
\begin{figure}[htbp] 
   \centering
   \includegraphics[width=\columnwidth]{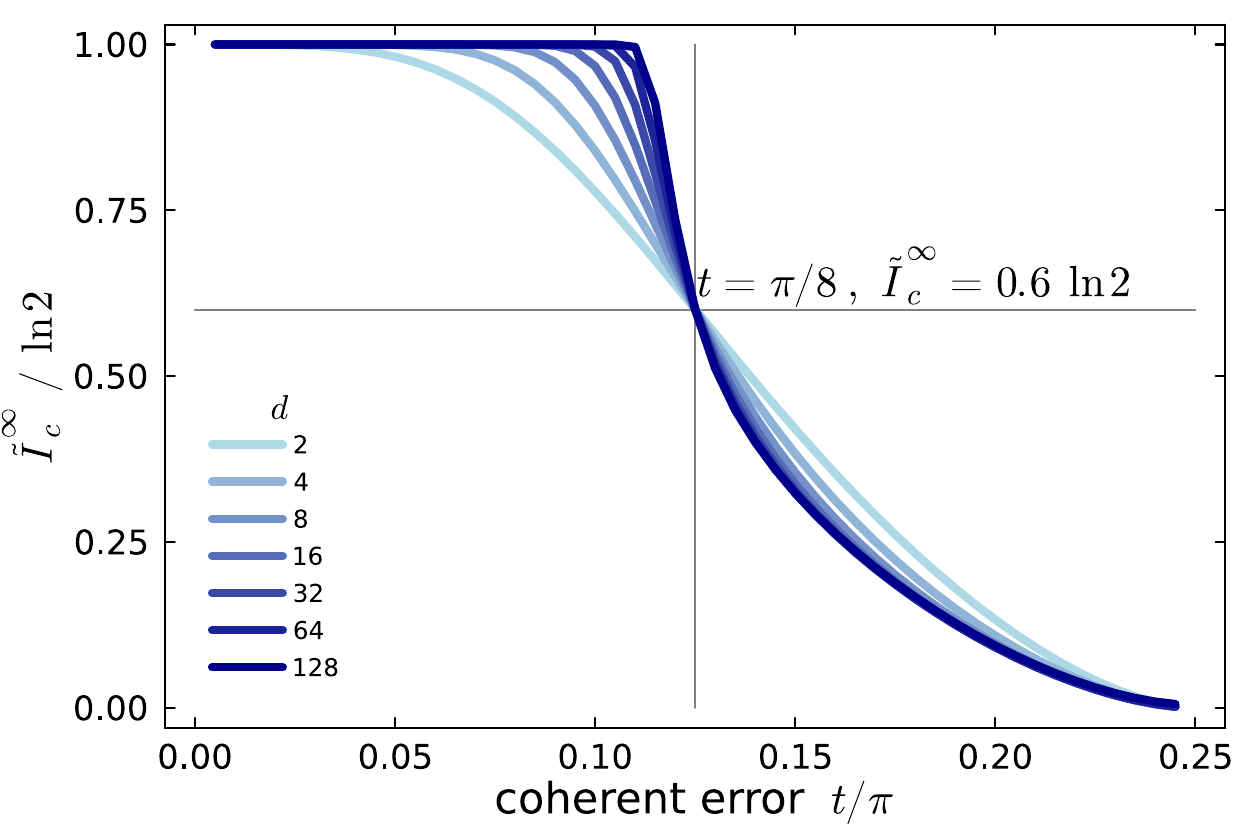} 
   \caption{{\bf Post-selection ($\mathbf{\infty}$-replica) model along self-dual line $\mathbf{X+Z}$}. 
   The gray lines indicate $(t=\pi/8, I_c^{\infty}=0.6\ln 2)$. 
   Note the clear level crossing for different code distances in the plotted coherent information occurring 
   for the Kosterlitz-Thouless transition at the 4-state Potts point $t_c=\pi/8$. 
   While Eq.~\eqref{eq:purity} at $n\to \infty$ limit reduces to $I_c^\infty = -\ln \frac{1+\kappa}{2}$, 
   we here show the von Neumann version $\tilde{I}_c^\infty = -\frac{1+\kappa}{2}\ln \frac{1+\kappa}{2}-\frac{1-\kappa}{2}\ln \frac{1-\kappa}{2}$, 
   where the two are in one-to-one correspondence and behave similarly. 
   MPS cutoff $10^{-10}$ and $\chi=1024$. 
   }
   \label{fig:postselectselfdual}
\end{figure}

The local transfer matrix describes a clean XXZ spin chain
\begin{equation}
\begin{split}
T^{(\infty)}  &\propto \left(\frac{1-Z_{j}Z_{j+1}}{2}\right)+ \frac{\sin(2t)}{\sqrt{2}} \left(\frac{1+\vec{\sigma}_{j}\cdot \vec{\sigma}_{j+1}}{2}\right)\ .
\end{split}
\end{equation}
The critical point lies at $t_c=\pi/8$ which opens a continuously varying critical line for $t\geq \pi/8$~\cite{Zhu19}. While the Kosterlitz-Thouless transition usually does not show clear level crossing for the spin wave stiffness, we find that the von Neumann coherent information of the post-selected case does show a perfect crossing here (Fig.~\ref{fig:postselectselfdual}), which means $-\langle Z_0Z_{2d+1}\rangle=\langle \prod_{j=1}^{2d}Z_j\rangle$ has zero scaling dimension. $I_c$ approaches $1$ exponentially fast for $t<t_c=\pi/8$ below the threshold, but converges quickly to a continuously varying finite constant for $t>t_c$. 

\section{Supplementary derivations}


\subsection{Teleportation circuit block}
To understand the teleportation circuit, first consider a two-qubit $R_{ZZ}$ rotation gate followed by measurement in $X$ basis
\begin{equation*}
\includegraphics[width=.75\columnwidth]{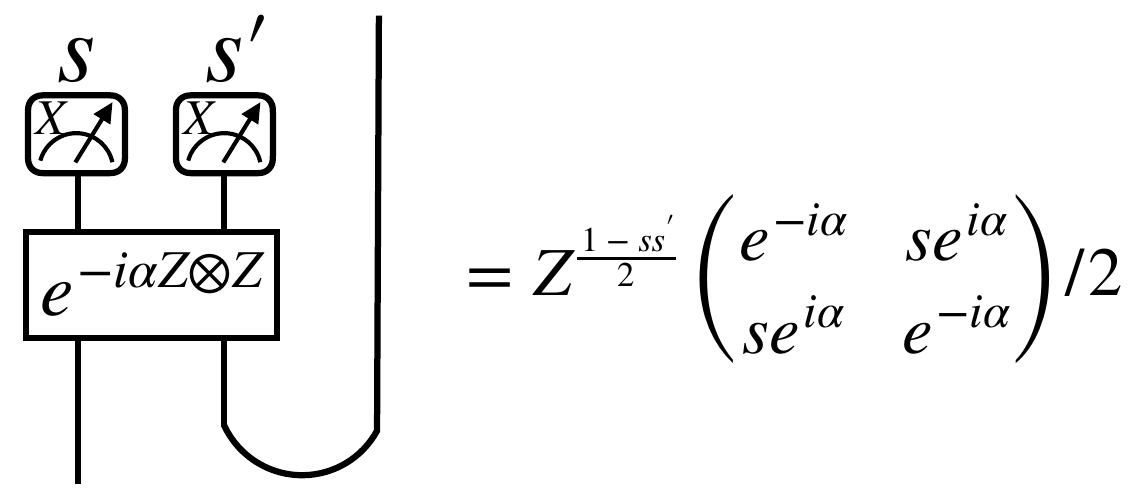} \ ,
\end{equation*}
where the $R_{ZZ}$ and the $X$-measurement outcomes all play the role of attaching a phase factor to the four possible input and output states, framed in a non-unitary 2-by-2 matrix. This matrix transports the input qubit wire from the bottom left to the output qubit wire at the top right. 
Next we perform a Hadamard rotation that swaps $X\leftrightarrow Z$:
\begin{equation*}
\includegraphics[width=.7\columnwidth]{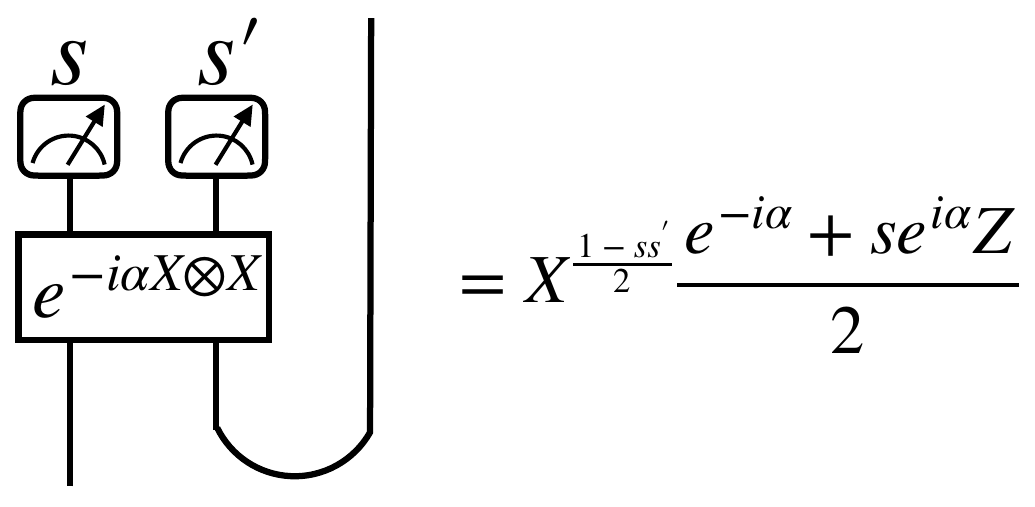} \ .
\end{equation*}
Finally, we perform the simple on-site correction conditioned on the measurement outcomes (without need for decoder)
\begin{equation}
\includegraphics[width=.99\columnwidth]{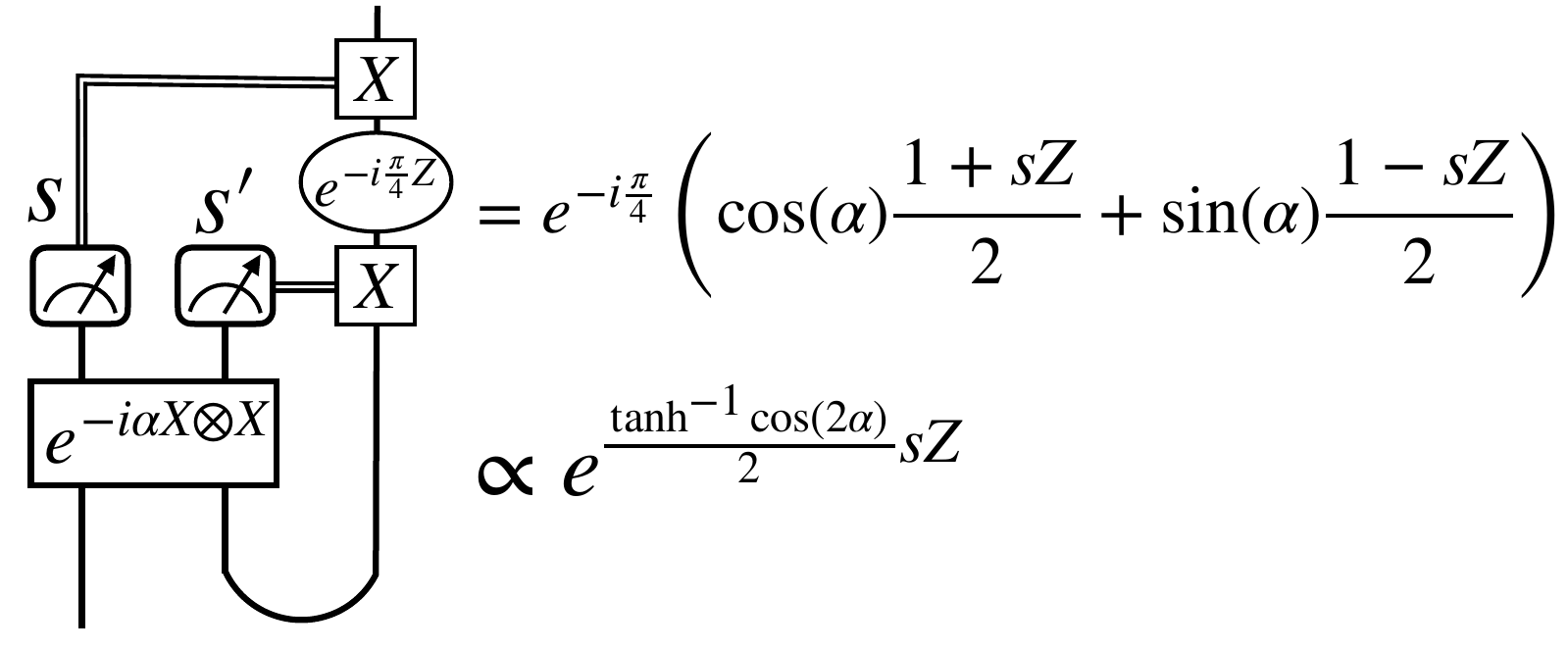} \ ,
\label{eq:circuit}
\end{equation}
that leads to a real non-unitary matrix as purely imaginary time evolution. This is equivalent to the weak measurement in the $Z$ basis~\cite{NishimoriCat}
\begin{equation*}
\includegraphics[width=.5\columnwidth]{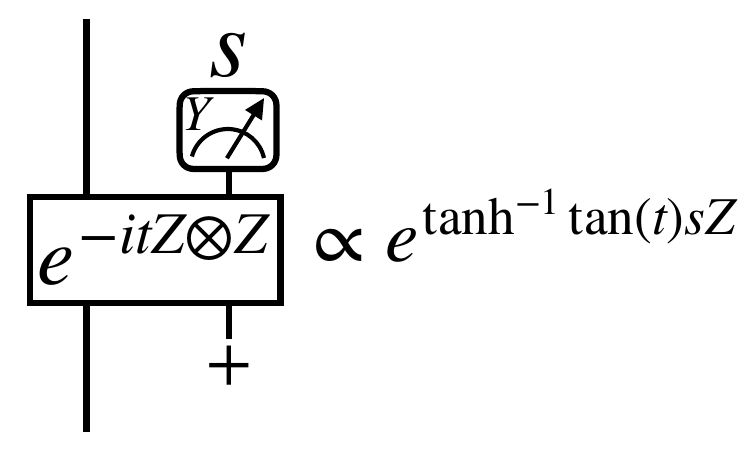} \ ,
\end{equation*}
which can be summarized as normalized Kraus operator
$\exp(\frac{\beta}{2} sZ)/\sqrt{2\cosh\beta}$
where $\tanh(\beta)=\cos(2\alpha)=\sin(2t)$ stands for the effective measurement strength. 
Here we define $t=\pi/4-\alpha$ to characterize the deviation from the perfect teleportation protocol, because $\alpha=\pi/4$ corresponds to a perfect Bell pair measurement that teleports the input qubit state into the output qubit without decoherence, which is consistent with $t=0$ zero measurement strength that decoheres the state.  
Note that the randomness of $s'$ is simply corrected and drops out from the formula, but the randomness of $s$ cannot be fully undone. 
As the Ising evolution gate can be decomposed into a single-body rotation sandwiched by CNOT gates:
\begin{equation*}
\includegraphics[width=.5\columnwidth]{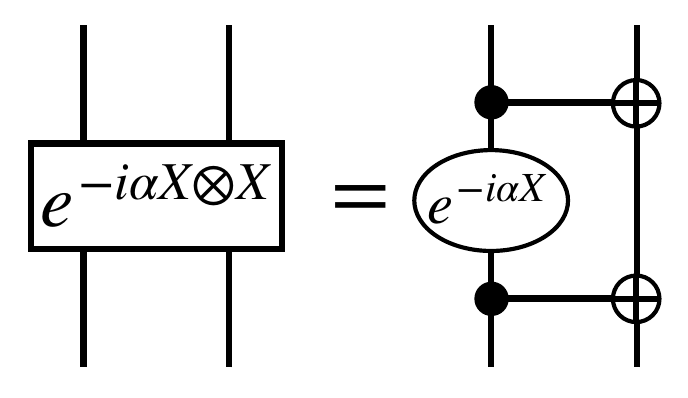} \ ,
\end{equation*}
the 2-body coherent error $t$ can result from the single-body rotation.
Last but not least, we make a comment about the more general situation where the Bell pairs are created imperfectly by an imperfect entangling gate $R_{XX}$ up to angle $\alpha'$:
\begin{equation*}
\includegraphics[width=.6\columnwidth]{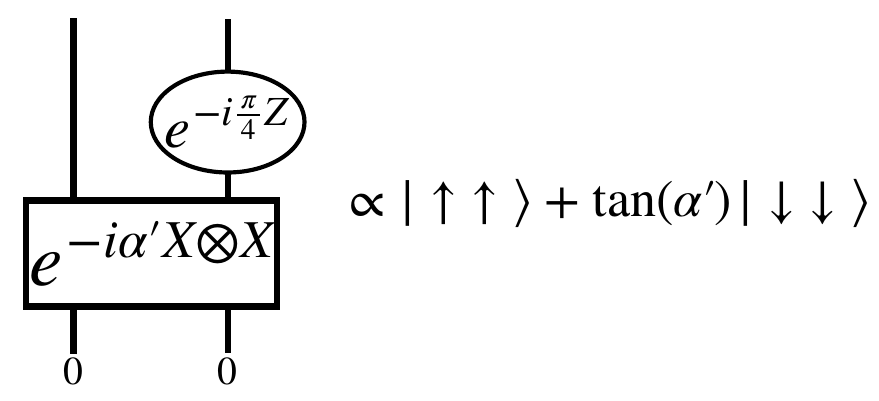} \ ,
\end{equation*}
which state is akin to a non-interacting thermofield double state at finite temperature $-1/\ln\tan(\alpha')$. Combining the joint actions of imperfect Bell preparation and imperfect Bell measurement, the post-teleportation state suffers from the following effective weak measurement gate operator:
\begin{equation}
\includegraphics[width=.99\columnwidth]{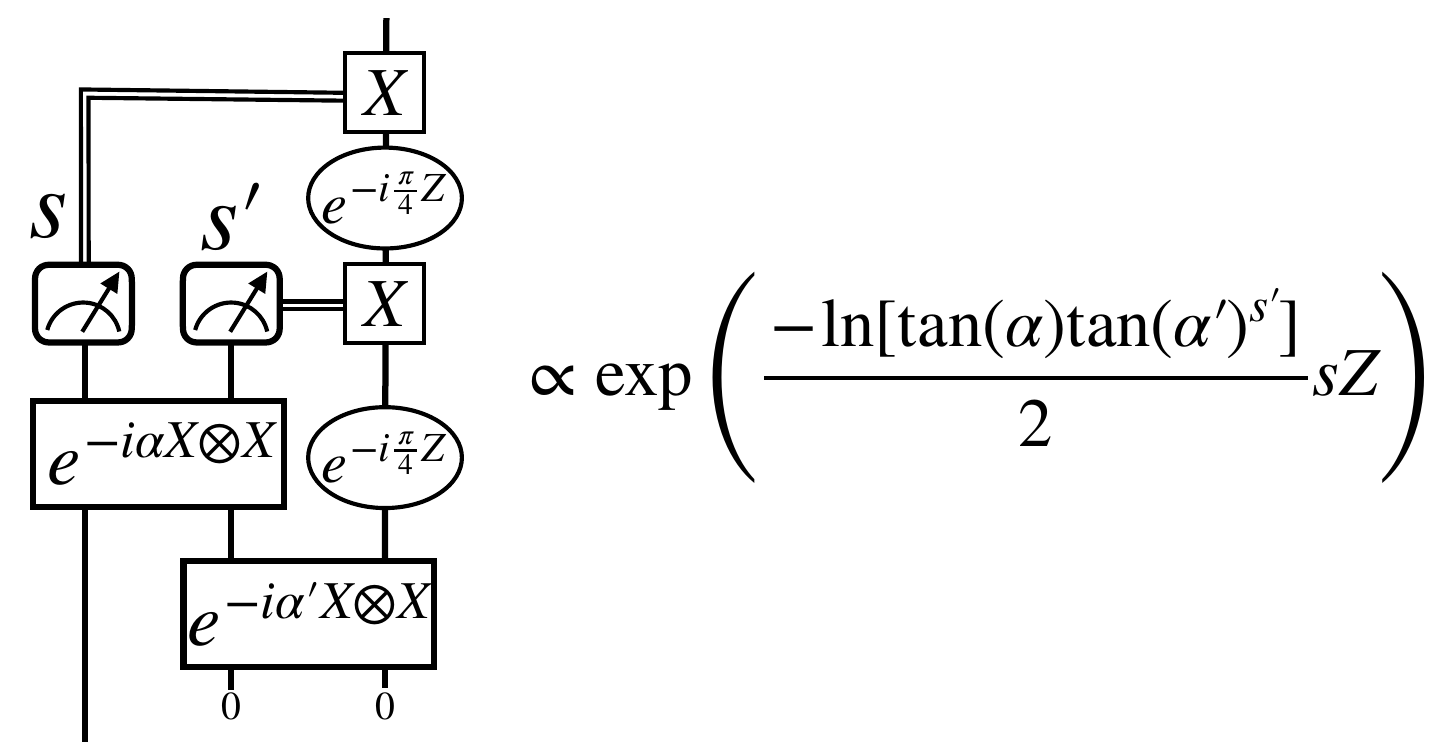} \ ,
\end{equation}
which is akin to what we discuss in the main text with minor adaptation. One can verify that (i) when $\alpha'=\pi/4$ for perfect Bell pairs, it recovers Eq.~\eqref{eq:circuit}; (ii) when $\alpha=\pi/4$ for perfect Bell measurement, it also recovers Eq.~\eqref{eq:circuit} by replacing $s\to ss'$ and $\alpha\to\alpha'$; (iii) when both Bell pair preparation and Bell pair measurement are imperfect, the random disorder now depends on both $s$ and $s'$ quaternary rather than binary disorder, which details we leave to future study. 

For completeness, here we also derive a building block for the quantum state transfer without using Bell pair, but simply transferring a state from Alice to Bob via direct entangling and measuring Alice, as follows:
\begin{equation}
\includegraphics[width=.99\columnwidth]{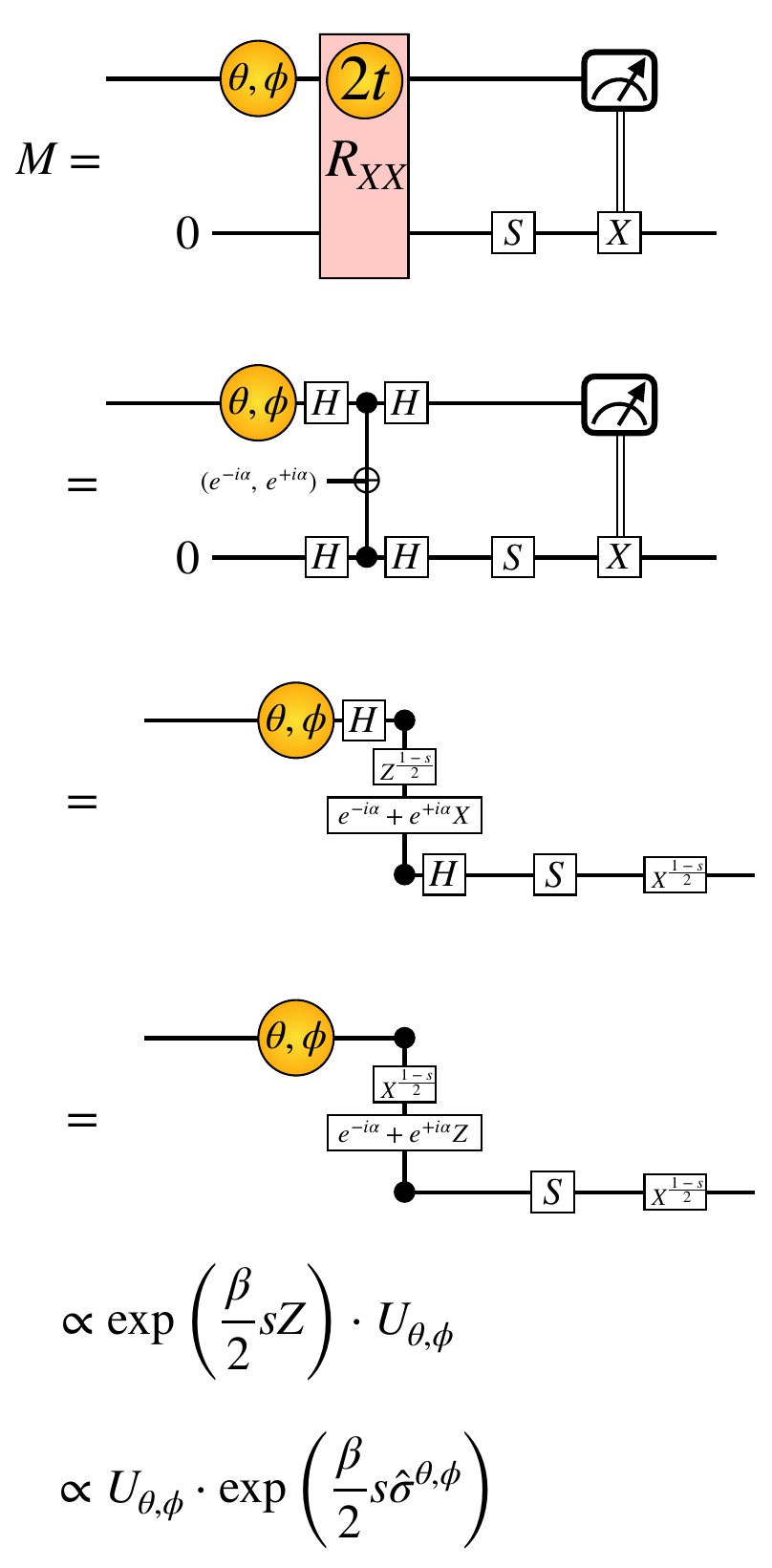} \ .
\end{equation}
Here we initiate Bob in a $0$ state, and rotate Alice by a unitary transformation $U_{\theta,\phi}$ that rotates $U_{\theta,\phi}^\dag Z U_{\theta,\phi} = \hat{\sigma}^{\theta,\phi}$. Then we entangle them with an $R_{XX}$ gate over an angle $\pi/4-t$, where $t\in[0,\pi/4]$ specifies the strength of coherent error. Afterwards we projectively measure out Alice and perform an onsite correction for Bob, by an $S=\text{diag}(1,i)\propto e^{-i\frac{\pi}{4}Z}$ gate and a conditional $X$ gate. In total, the state in Alice is transferred to Bob experiencing an effective non-unitary gate, which can be interpreted as a weak measurement (due to coherent error) with $\beta$ being the measurement strength. As a result, this state transfer protocol also shares exactly the same phase diagram as we chart out in this paper. 


\subsection{Tensor network representation}
\begin{figure}[tb] 
   \centering
   \includegraphics[width=.9\columnwidth]{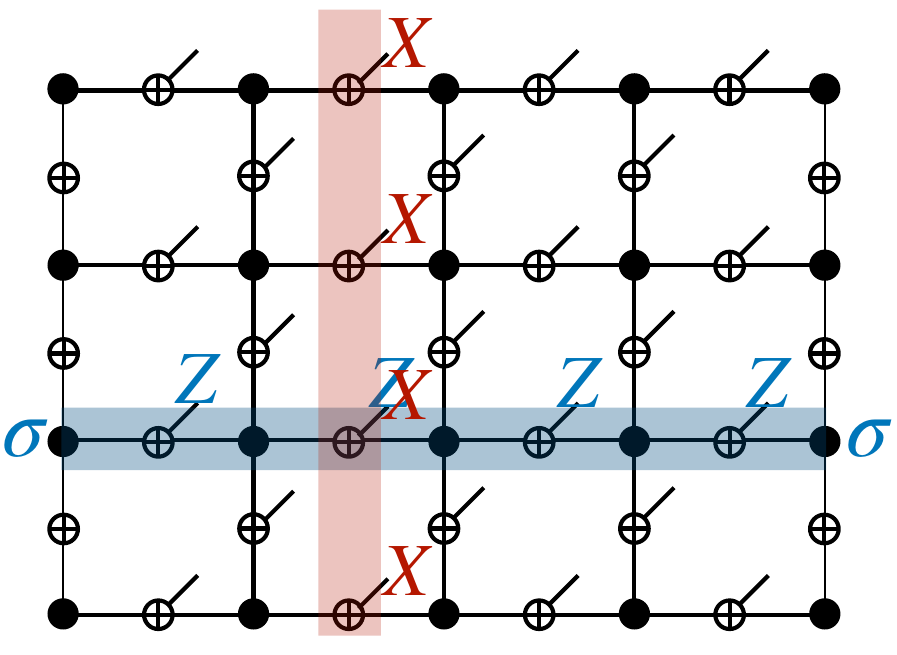} 
   \caption{{\bf Surface code PEPS}. 
   	Each solid dot denotes the diagonal delta tensor: $T_{ijkl} = 1$ iff $i=j=k=l$ (which represents a virtual GHZ state); and each cross node denotes the off-diagonal tensor: $T_{ijk} = 1$ iff $i+j+k\mod2=0$ (which represents a virtual GHZ state in $X$ basis: $\ket{+++}+\ket{---}$). 
	The left and right rough boundaries captured by removing the dangling physical leg, such that the boundary forms a ferromagnetic GHZ chain, consistent with $e$ condensation. The $Z$ string acting on the physical legs can be {\it pulled through} to the virtual leg reduced into a two-point $\sigma-\sigma$ operator connecting the two boundaries. Thus the logical qubit of the surface code that is changed by $Z$-string corresponds to the total Ising symmetry charge of the left and right GHZ chains.
	The top and bottom boundaries are smooth boundary, where $m$ particle is condensed. 
	}
   \label{fig:PEPS}
\end{figure}

The surface code can be written as a projected entangled pair state as illustarted in Fig.~\ref{fig:PEPS}. 
By stacking two layers of PEPS together and tracing out the physical legs, we obtain a 2D classical tensor network composed of delta tensor at each vertex, joined by bond matrices as follows:
First, we write down and simplify the doubled Kraus operator matrix:
\begin{equation}
\begin{split}
M_s^\dag M_s&=e^{\beta s (Z \cos\theta + \sin\theta(\cos\phi X+\sin\phi Y))} /(2\cosh\beta)\\
&= \frac{1}{2\cosh J}\left(
\begin{matrix}
e^{sJ} & se^{-2K} e^{-i\phi} \\
s e^{-2K} e^{+i\phi} & e^{-sJ}\\
\end{matrix}
\right) \ ,
\end{split} \ ,
\end{equation}
where we introduce $J$, $K$ as functions of $t$ and $\theta$:
\begin{equation}
\tanh(J)  = \sin(2t) \cos(\theta) \ , \quad 
e^{-2K}= \sinh(J) \tan(\theta) \ .
\end{equation}
Then we introduce the wave function ket and bra:
\begin{equation}
\ket{\psi} = \sum_{\{\sigma=\pm1\}}\bigotimes_{ij}\ket{Z_{ij}=\sigma_i\sigma_j},
\quad
\bra{\psi} = \sum_{\{\tau=\pm1\}}\bigotimes_{ij}\bra{Z_{ij} = \tau_i \tau_j},
\end{equation}
expressed as a dual paramagnet of classical spins $\sigma$ and $\tau$. 
As a result,
\begin{equation}
\bra{\psi} M_s^\dag M_s\ket{\psi} \propto \sum_{\sigma,\tau} \exp(-\sum_{\langle ij\rangle} E_{ij}) \ ,
\end{equation}
up to an $\mathbf{s}$ independent constant prefactor, 
where $E_{ij}$ is the generalized Ashkin-Teller model shown in the main text.

To further reduce the bilayer model into a single layer vertex model / XXZ chain, there are two alternative ways. One way is to perform duality for one layer only, resulting in the 8-vertex model discussed in the main text. 
An alternative way is to take a slice of the network as a quantum transfer matrix for $2(d+1)$ spins subjected to Ashkin-Teller interactions, which can be rewritten in terms of XXZ interactions under the mapping
\begin{equation}
\begin{split}
&\sigma_j^z\sigma_{j+1}^z = X_{2j} X_{2j+1} , \quad \sigma_j^x = Y_{2j-1} Y_{2j} ,\\
& \tau_j^z\tau_{j+1}^z = Y_{2j} Y_{2j+1}, \quad \tau_j^x = X_{2j-1} X_{2j} \ ,
\end{split}
\label{eq:mapAT2XXZ}
\end{equation}
which preserves all Pauli commutation relations. Consequently, the interaction for the horizontal gates (see Fig.~\ref{fig:2rep} of the main text) is

\begin{equation}
\begin{split}
&-E_{2j,2j+1} \\
=&\frac{Js}{2}(X_{2j}X_{2j+1}+Y_{2j}Y_{2j+1}) +i\frac{\phi}{2} (X_{2j}X_{2j+1}-Y_{2j}Y_{2j+1})\\
&- (K+i\frac{\pi}{2}\frac{1-s}{2})(Z_{2j}Z_{2j+1}+1) \ ,\\
=&Js (c^\dagger_{2j} c_{2j+1} + h.c.) + i \phi (c_{2j} c_{2j+1} + h.c.) \\
&- 2(K +i\frac{\pi}{2}\frac{1-s}{2})(2n_{2j}n_{2j+1}-n_{2j} - n_{2j+1}+1)
\end{split}
\end{equation}
which is an antiferromagnetic anisotropic Heisenberg chain with PT symmetric non-Hermitian interactions, that is mapped to a complex fermion chain $c_j$ with non-hermitian pairing term. Here $n_j=c_j^\dag c_j$ denotes fermion density. The matrix elements of $\exp(-E)$ in the $Z$ basis yields the Boltzmann weight for the corresponding vertex configuration, as shown in the right panel of Fig.~\ref{fig:2rep} in the main text. The vertical gate can be obtained by rotating the horizontal gate employing the duality of the model.

\subsection{Coherent information cast in the classical model}
For simplicity of explanation, let us first consider $\theta=0,\phi=0$ i.e.\ the $Z$ direction, where $P_{--}=P_{++}$, and the classical statistical model is a random bond Ising model. 
The density matrix $\rho_R$ describing the reference qubit is purely diagonal in $Z$ basis and classic: by adapting Eq. (5) in our main text from Pauli $X$ basis to $Z$ basis we have $\rho_R=\text{diag}(P_{+-}/P_{++}, P_{+-}/P_{++})$, where according to our dictionary (Table I in main text) $P_{+-}/P_{++} = \langle \sigma_0\sigma_d\rangle$, which is the two-point correlation between the left boundary and the right boundary. 
The classical model is placed on the following ``fixed-boundary" geometry (inherited from the surface code boundary condition)
\begin{equation}
\includegraphics[width=\columnwidth]{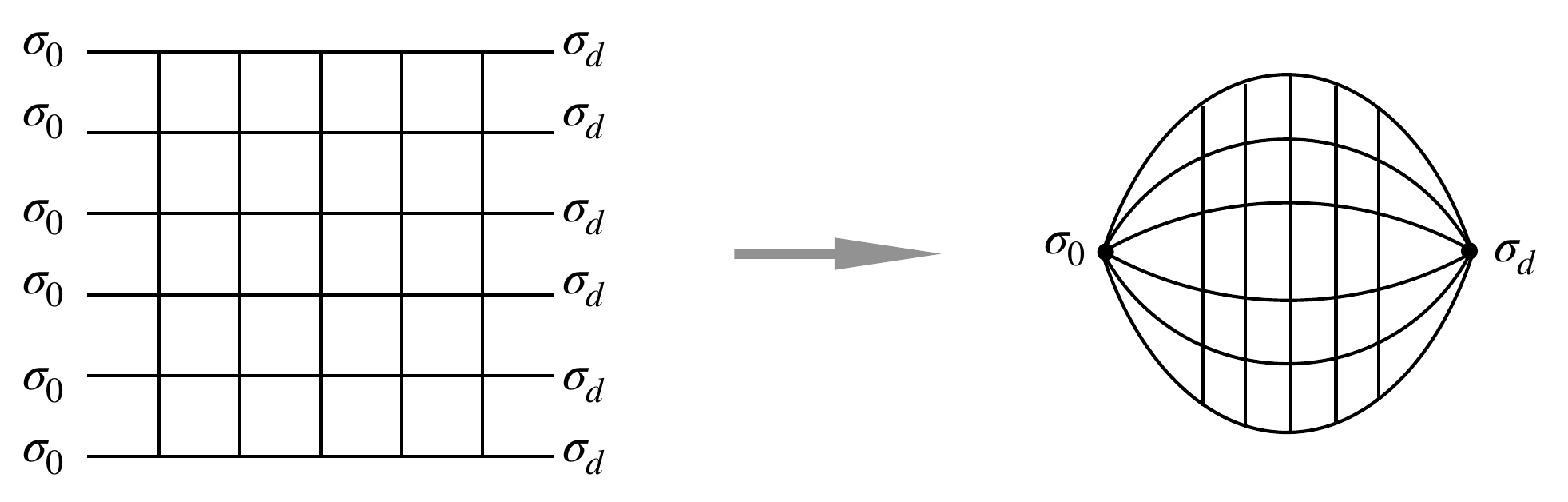} \ ,
\end{equation}
where all spins on the left/right boundary are fixed to identical values $\sigma_0$ or $\sigma_d$, respectively. In the original surface code, the "rough" boundary (illustrated above on the left hand side) condenses the $e$ particle on the left or right thereby effectively glueing the whole boundary into one site (as shown on the right hand side above). 
The parity of the left and right boundary sites can be used to define one bit $\kappa = \sigma_0 \sigma_d$, whose distribution function is $(1\pm \langle \kappa\rangle)/2$. The classical entropy of this ``boundary bit"
\begin{equation}
S = -\frac{1+\langle \kappa\rangle}{2}\ln \frac{1+\langle \kappa\rangle}{2} -\frac{1-\langle \kappa\rangle}{2}\ln \frac{1-\langle \kappa\rangle}{2} \ ,
\end{equation}
exactly equals the quantum coherent information of a given post-measurement state with randomness. The coherent information of the mixed state is the random average of this entropy according to the Born's probability. 
Importantly, this entropy should not be confused with the boundary entropy of the boundary CFT. 

When deviating from the $Z$ direction, the RBIM turns into a random Ashkin-Teller model in a similar geometry, where the coherent information is again determined by the correlation between the leftmost and rightmost sites in this special boundary condition. With two layers of classical spins, in general we can define a boundary quantum bit subject to the following ``magnetic field" $\vec{\kappa}$:
\begin{equation*}
\begin{split}
&\kappa_x = \frac{P_{++}-P_{--}}{P_{++}+P_{--}}=\frac{1-\langle \sigma_0\tau_0\sigma_d\tau_d\rangle}{1+\langle \sigma_0\tau_0\sigma_d\tau_d\rangle} \ , \\
&\kappa_z+i\kappa_y = \frac{2P_{+-}}{P_{++}+P_{--}}=\frac{2\langle \sigma_0\sigma_d\rangle}{1+\langle \sigma_0\tau_0\sigma_d\tau_d\rangle} \ . \\
\end{split}
\end{equation*}
Then the density matrix eigenvalues of this boundary quantum bit is $(1\pm \norm{\vec{\kappa}})/2$, whose corresponding entropy is again equivalent to the coherent information of the quantum system.


\subsection{Rényi coherent information and replica symmetry}
The $n$-th order Rényi coherent information in the main text can be easily derived by tracing out $B$ resulting in the reduced density matrices for $RA$ and $A$:
\begin{equation}
\begin{split}
\rho_{RA} = \sum_\mathbf{s} P(\mathbf{s}) \cdot \rho_R(\mathbf{s}) \otimes \ketbra{\mathbf{s}} \ ,\ 
\rho_A = \sum_\mathbf{s} P(\mathbf{s}) \ketbra{\mathbf{s}} \ ,
\end{split}
\end{equation}
whose conditional $n$-th order Rényi entropy is determined by the $n$-replica of the classical random Ashkin-Teller model composed of $2n$ layers of Ising spins, denoted by a flavour index $\alpha=1,\cdots,2n$. 
For the asymptotic limit $\phi=0,\ t\to\pi/4$ for any $\theta$ (the circular edge of our phase diagram in Fig.~1 c), the Ashkin-Teller coupling between every two layers vanishes $K\to0$, leaving only by the phase factor $s^{\frac{1-\sigma_i\sigma_j\tau_i\tau_j}{2}}$. Consequently, there is an emergent $S_{2n}$ permutation symmetry in the flavor space:
\begin{equation}
\begin{split}
&\sum_\mathbf{s}P(\mathbf{s})^n =\\
& \sum_{\sigma}\sum_\mathbf{s} \exp{\sum_{\langle ij\rangle}
\frac{J}{2} s_{ij} \sum_{\alpha=1}^{2n} \sigma_i^\alpha\sigma_j^\alpha
+i\pi \frac{1-s_{ij}}{2} \frac{1-\prod_{\alpha=1}^{2n} \sigma_i^\alpha\sigma_j^\alpha}{2}
} \ ,
\end{split}
\end{equation}
where the $2n$ Ising layers are only coupled by an $S_{2n}$ symmetric interaction. The $S_{2n}$ symmetric model is self-dual at $\theta=\pi/4$. 
For example, the 2-replica model has emergent $S_4$ permutation symmetry and self-duality at $(t=\pi/4,\ \theta=\pi/4,\ \phi=0)$, which is expected to be described by the 4-state Potts conformal field theory, the same as the post-selection $\infty$-replica at $(t=\pi/8,\ \theta=\pi/4,\ \phi=0)$.

\section{Experiment and quantum error correction}

We believe that our protocol is amenable to a direct implementation in reconfigurable Rydberg atom arrays~\cite{Browaeys2020,Lukin2021atom256,Lukin2023logical}. 
In this appendix, we address how to decode the teleportation transition in such experimental implementations. 

As the logical qubit cannot be cloned, the basic idea is to investigate whether it falls into Alice's hand or Bob's hand. The teleportation succeeds if Bob has the key, and fails if Alice has it. This teleportation of a single logical qubit is in one-to-one correspondence with teleporting the many-body $Z_2$ topological order, quantified by the topological degeneracy. 
When there is no other eavesdropper, detecting Alice and detecting Bob shall show the same transition. The difference between Alice and Bob, however, is that Alice has a classical ensemble of measurement bit strings $\{\mathbf{s}\}$, while Bob has an ensemble of quantum wave functions $\{ \psi(\mathbf{s})\}$ that come along with the message $\mathbf{s}$ shared from Alice.

\subsection{Decoding Alice}
Here we investigate whether Alice's measurement outcomes are sufficient to infer the information stored in the reference qubit~\cite{Gullans20scalabledecoder}. 
To make the teleportation transition observable we will follow a three-step protocol:

First, we projectively measure the reference qubit $R$ in the $Z$ basis together with projective measurements of Alice's qubits $A$ 
in every run of the experiment. 
As a result, we get an ensemble of bit-strings capturing $\mathbf{s}$ for Alice qubits and $k=0(1)$ for the reference qubit. 

Second, to see the statistical correlation between $R$ and $A$, we use a standard diagnostic, the classical Shannon relative entropy. This quantity is equivalent to the entropy of $R$ averaged over the $A$ measurement outcomes:
\begin{equation}
I_c^{z}
=
[S_R^z(\mathbf{s})]
=[\langle \ln \frac{P(\mathbf{s})}{P_k (\mathbf{s})}\rangle ]
\ ,
\end{equation}
which is to take $\ln  \frac{P(\mathbf{s})}{P_k (\mathbf{s})}$ (calculated by a classical computer) being averaged over all the experimentally obtained bit-strings. 
Here for the Shannon entropy, we need a classical decoder (running the same tensor network calculation as we do here) that computes the non-linear part $\ln P_{k}(\mathbf{s})$, corresponding to each important sample bit-string obtained from the experiment. 

Finally, we make use of the symmetry for the protocol to reconstruct the full coherent quantum information: 
\begin{enumerate}
\item $\theta=0$: the logical space is compressed only in the $X$ direction, and thus the $Z$-basis measurement directly leads to the full coherent information: $I_c=I_c^{z}$.

\item $\theta=\pi/4$: the model maintains self-duality, such that the $X$ contribution is equal to the $Z$ contribution. And thus when we consider only the diagonal entry, the density eigenvalues of the reference qubit (for each $\mathbf{s}$) reduces from $(1\pm \kappa)/2 \mapsto (1\pm \kappa/\sqrt{2})/2$, which would rescale the coherent information: $I_c \mapsto I_c^{z}$. We can revert this map to obtain the full coherent information from each bit-string (for only $\theta=\pi/4$).

\end{enumerate}

\subsection{Decoding Bob}

Besides guiding the simple local correction as in the standard {\it few-body} teleportation experiment, Alice's measurement outcome $\mathbf{s}$ can additionally serve as {\it many-body} syndrome. 
In the active teleportation protocol, where Bob has access to these syndromes, he can deduce the underlying errors by running our classical calculation to contract out the tensor network of Fig.~3 in main text, which works for general angles. 

Let us comment on the active teleportation versus passive teleportation protocol in the following. 
For the active teleportation along $\theta=0$ ($\phi=0$) the quantum wave function is mapped to the RBIM from high to low temperature, as visualized as in Fig.~4b ``hot" to ``cool" regimes. Here ``temperature" describes the uncertainty of the measurement outcomes. Consequently, it exhibits a finite threshold at the Nishimori critical point at $t_c\approx 0.143\pi$. The passive teleportation protocol, in contrast, does not use a classical decoder on the way to process the information, but instead mixes the pure states into a noisy surface code mixed state: $\mathcal{N}(\rho) = \cos^2(t) \rho + \sin^2(t) Z \rho Z$, which can be mapped to the RBIM from low to high temperatures. In this case, ``temperature" describes the fluctuations induced by noise. Consequently, its optimal threshold lies at the opposite location $t_c\approx 0.107\pi = \pi/4-0.143\pi$.

\end{document}